\documentclass{article}
\usepackage{amsfonts, amssymb, amsmath, chicago, rotating, latexsym}
\newtheorem{theorem}{Theorem}[section]

\newtheorem{lemma}{Lemma}[section]

\newtheorem{prop}{Proposition}[section]

\begin{document}

\title{{\LARGE Threshold estimation based on a $P$-value framework}}
\author{Bodhisattva Sen, Moulinath Banerjee and George Michailidis \\ {\it Columbia University}, {\it University of Michigan} and {\it University of Michigan}}
\maketitle

\begin{abstract}
We use $p$--values as a discrepancy criterion for identifying the threshold value at which a regression function takes off from its baseline value -- a problem that is motivated by applications in omics experiments, systems engineering, pharmacological dose-response studies and astronomy. In this paper, we study the problem in a controlled sampling setting, where multiple responses, discrete or continuous, can be obtained at a number of different covariate-levels. Our procedure involves testing the hypothesis that the regression function is at its baseline at each covariate value using the sampled responses at that value and then computing the $p$--value of the test. An estimate of the threshold is provided by fitting a stump, i.e., a piecewise constant function with a single jump discontinuity, to the observed $p$--values, since the corresponding $p$--values behave in markedly different ways on different sides of the threshold. The estimate is shown to be consistent, as both the number of covariate values and the number of responses sampled at each value become large, and its finite sample properties are studied through an extensive simulation study. Our approach is computationally simple and can also be used to estimate the baseline value of the regression function. The procedure is illustrated on two motivating real data applications. Extensions to multiple thresholds are also briefly investigated.
\end{abstract}

\section{Introduction and Problem Formulation}
\begin{figure}
\centering
\resizebox{5.5in}{2in}{\includegraphics{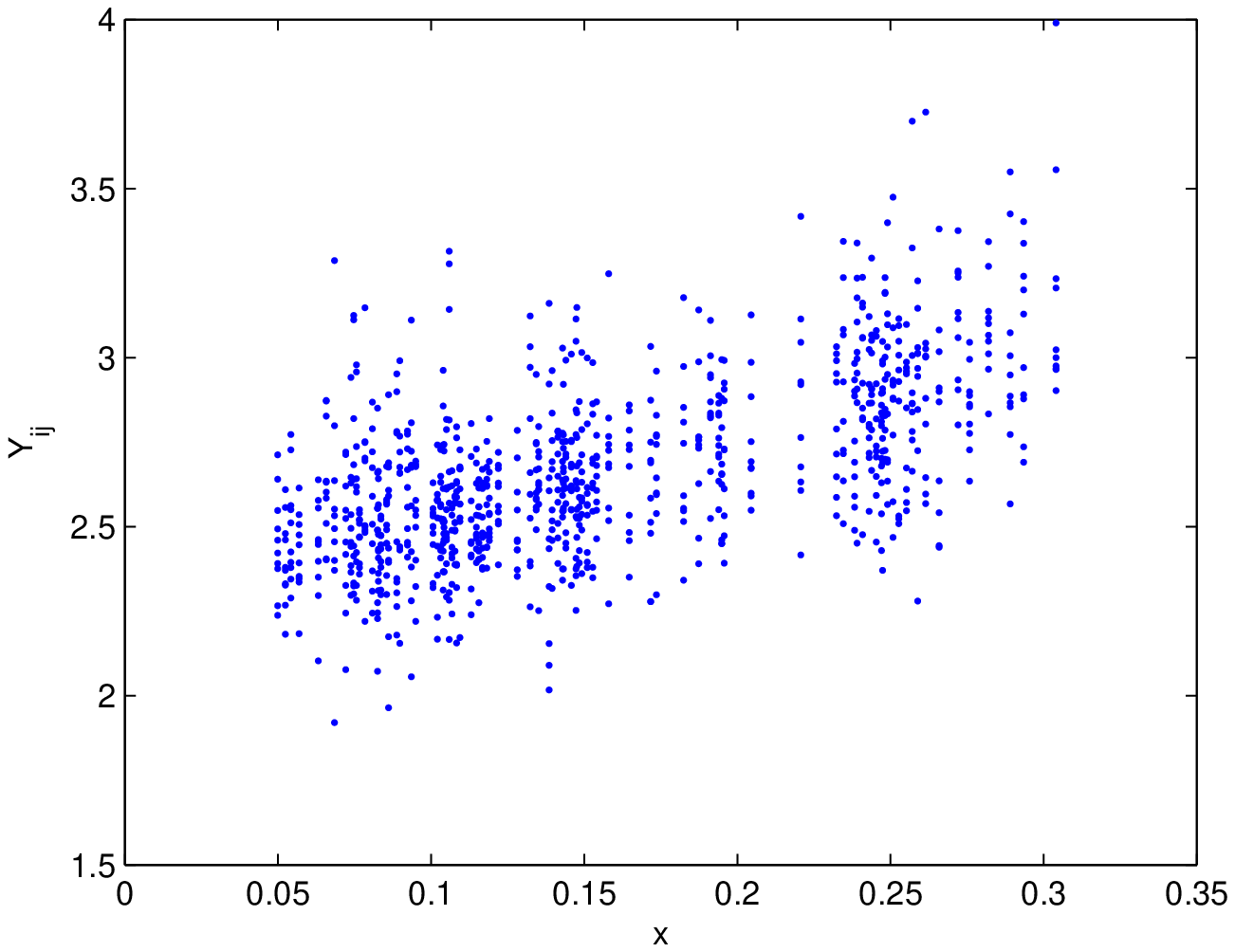}  \includegraphics[scale=.74]{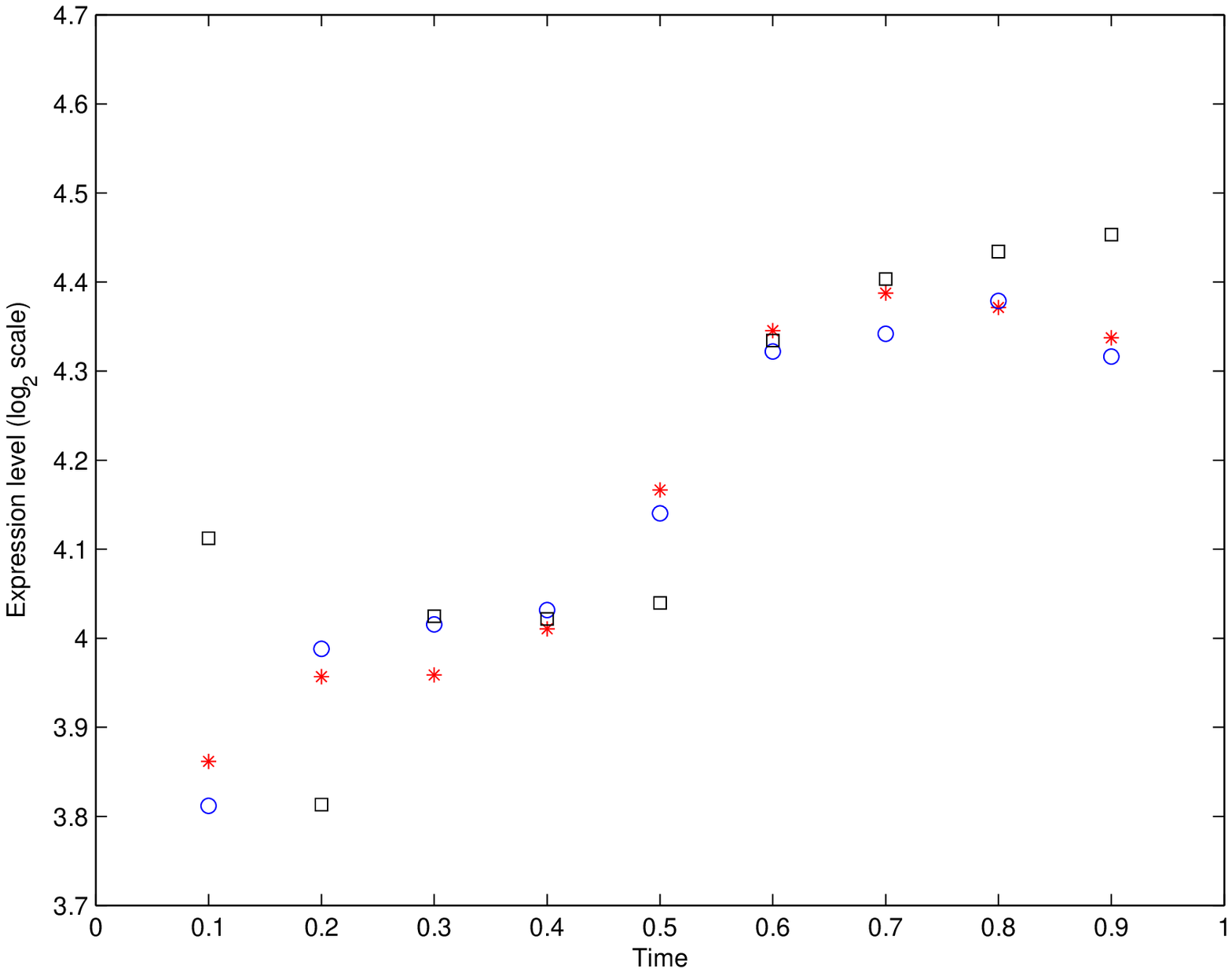} }
\caption{Left panel: Data of delay versus loading from a complex queueing system.
Right panel: Gene expression levels over time. \label{full-data} }
\end{figure}

In a number of applications, the data follow a regression model where the regression function $\mu$ is constant at its baseline value $\tau_0$ up to a certain covariate threshold $d^0$ and stays above $\tau_0$ at higher covariate levels. For example, consider the data in the left panel of Figure \ref{full-data} that depict the average delay of customers as a function of the loading of a complex queueing system (more details about the system are given in Section \ref{data-application}). It can be seen that for small loadings the delay is rather small, while a clear positive trend emerges for larger loadings. The system's operator is interested in identifying this threshold with high precision, as well as the level of the baseline $\tau_0$, since such knowledge specifies an operational regime of low average delay, whose value can be announced to potential customers. An example from a different scientific domain is shown in the right panel of Figure \ref{full-data}. It depicts the expression levels of a gene over time obtained from multiple cell-lines, where again the function stays at its baseline level up to some time, then rises sharply and subsequently flattens out (more details about the underlying experiment are given in Section \ref{data-application}). This problem requires procedures that can handle multiple change-points of the regression function, namely where it deviates from the baseline value and also where it starts flattening out towards the end of the range of the covariate. These thresholds are of interest as they represent important stages of progression of the cell from normalcy to malignancy.
\newline
\newline
Problems with identical structure also arise in pharmacological dose-response studies, where $\mu(x)$ provides information about reaction to dose--level $x$ and is typically at the baseline value up to a certain dose, referred to in the literature as the minimum effective dose (MED); see Chen and Chang (2007) and Tamhane and Logan (2002) and the references therein. Similar problems arise in toxicological applications; see, for example, Cox (1987), who uses parametrically specified threshold models. Yet another field of application is astronomy; one of particular interest to the authors deals with estimating the ``tidal'' radius of a dwarf spheroidal galaxy (see Sen et al. (2009)). The mean velocity of the stars along the major axis of the dwarf spheroidal galaxy, as a function of the distance from the center, is assumed to be constant to the left of the ``tidal'' radius (threshold parameter), and takes off from this baseline value, due to interactions with the gravitational field of the Milky Way, as we move to the right of the threshold. Further applications and extensions are discussed in Section \ref{sec:conclusions}. We also note that our current problem of interest is a special and important case of the more general question of identifying the region where a function (defined on a general Euclidean space) assumes an extremal (minimum/maximum) value.
\newline
\newline
Formally, we consider a non--negative regression function $\mu(x)$ on $[0,1]$ with the property that $\mu(x) = \tau_0$ for $x \leq d^0$ and $\mu(x) > \tau_0$ for $x > d^0$ for some $d^0 \in (0,1)$. As already mentioned, quantities of prime interest are $d^0$ and $\tau_0$ that need to be estimated from noisy data $\{Y_i,X_i\}_{i=1}^n$, with $X_i$'s assuming values in $[0,1]$ and $Y_i = \mu(X_i) + \epsilon_i$, where $\epsilon_i$ is a mean 0 error. We call $d^0$ the ``$\tau_0$ threshold'' of the function $\mu$.
In this generality, i.e., without any assumptions on the behavior of the function in a neighborhood of $d^0$, the estimation of the threshold $d^0$ is a hard problem and has not been extensively addressed in the literature. In the simplest possible setting of the problem posited, the regression function $\mu(x)$ has a jump discontinuity at $d^0$. In this case, $d^0$ corresponds to a change--point for $\mu$ and the problem reduces to estimating this change--point in a regression model. Such change--point models are very well studied; see, for example, Mueller (1992), Loader (1996), Mueller and Song (1997), Koul and Qian (2002), Lan, Banerjee and Michailidis (2009) and references therein.
\newline
\newline
The problem becomes significantly harder when $\mu$ is continuous at $d^0$; in particular, the smoother the regression function in a neighborhood of $d^0$, the more challenging the estimation. For example, if $d^0$ is a cusp of $\mu$ of some \emph{known} order $p$ (i.e., the first $p-1$ right derivatives of $\mu$ at $d^0$ equal 0 but the $p$-th does not, so that $d^0$ is a change--point in the $p$-th derivative), one can obtain nonparametric estimates for $d^0$ using either kernel based (Mueller (1992)) or wavelet based (Raimondo (1998)) methods. The convergence rate of such estimates decreases dramatically with $p$ (see Raimondo (1998)), even with parametric specifications of $\mu$ to the right of the unknown $d^0$ (Feder (1975)). In fact, if $\mu$ is infinitely differentiable at $d^0$, no estimate possibly converges at a polynomial rate, although we are not aware of any such concrete result in the literature. However, in most applications, the degree of differentiability of $\mu$ at $d^0$ \emph{will not be known}, which makes necessary the development of adaptive estimation procedures that do not require prior information about the smoothness of the underlying regression function.
\newline
\newline
In many applications of interest, $\mu$ is known to be continuously increasing, and a natural candidate for an adaptive procedure is isotonic regression; further, it automatically avoids the specification of bandwidths and equivalent smoothing parameters (see Robertson et. al. (1988) and Silvapulle and Sen (2005)). Applications of isotonic regression in calibration type problems involving thresholds are discussed, for example, in Osborne (1991), Gruet (1996) and Tang et al. (2010). It is known that the isotonic regression estimate $\hat{\mu}$ is piecewise constant, usually possessing a flat stretch of 0 for small values of the covariate. Thus, one can prescribe $\hat{d} \equiv  \inf\{x : \hat{\mu}(x) > 0\}$ as an estimate of $d^0$, but this, in most cases, severely underestimates the true threshold. It is possible to employ penalized isotonic estimates or replace the 0 in the definition of $\hat d$ by a positive sequence $\eta_n$ converging to 0 (at an appropriate rate), but hardly anything is known in the literature about the theoretical properties of such procedures. Moreover, in many applications the assumption of global monotonicity of $\mu$ may not hold.
\newline
\newline
This paper develops a novel approach for the {\em consistent estimation} of $d^0$ for situations where \emph{multiple observations can be sampled} at a given covariate value that (i) does not require knowledge of the smoothness of $\mu$ at $d^0$, (ii) does not require computing an explicit estimate of $\mu$ and (iii) is computationally simpler than most nonparametric procedures. Note that this multiple observations per ``dose'' setting is the scenario for both our data applications and also for most of the dose--response studies in pharmacological experiments.
\newline
\newline
Note that, in both the motivating applications introduced in the first paragraph, the experimenter can specify the values of the covariate (either the load of the queueing system or the time point when the cell-line is harvested) and subsequently obtain the corresponding sample responses (delays or expression levels). In general, these sampled responses are expensive to obtain. In the first example, this is the case since generating the responses involves a discrete event simulation, with the response at each loading obtained by averaging over a number of events (e.g., customers who have received complete service from the system). For large scale systems, this may require tens of thousands of such events, which may exceed the allotted budget of resources. An alternative strategy is to rely on responses obtained from a fairly ``small" set of events for each selected loading, which would lead to significantly more noisy observations, as seen in the left panel Figure \ref{full-data}. In the second example, carrying out the biological experiment is fairly costly primarily due to the labor involved (preparation of cell-lines, microarray hybridization and processing). The obtained data can be noisy due to the inherent biological variability of the cell-lines. Nevertheless, the goal is to identify the transition point(s) and the corresponding levels of the threshold from such noisy data. The developed nonparametric methodology allows us to resolve this issue in a satisfactory manner. Specifically, it relies on testing for the value of $\mu$ at design levels of the covariate. The obtained test statistics are then used to construct $p$--values which, under mild assumptions on $\mu$, behave in markedly different fashions on either side of the threshold $d^0$ and it is this discrepancy that is used to construct an estimate of $d^0$.
\newline
\newline
The remainder of the paper is organized as follows: in Section \ref{sec:p-value-proc}, the proposed procedure is introduced in the case of {\em known} $\tau_0$, its properties derived and some extensions discussed. The case of {\em unknown} $\tau_0$ is examined in Section \ref{sec:unknown-tau}. Section \ref{multiple-pts} briefly investigates the generalization to multiple change points. The performance of the procedure based on simulated data is studied in Section \ref{sec:simulations} where comparisons with other competing methods are also presented. The procedure is also illustrated on real data from the two motivating applications. Some concluding remarks are drawn in Section \ref{sec:conclusions}, while proofs of most technical results are given in the Appendix.
\section{The $p$--value procedure: the known $\tau_0$ case}\label{sec:p-value-proc}
\begin{figure}[!h]
\centering
\resizebox{3.5in}{2in}{\includegraphics{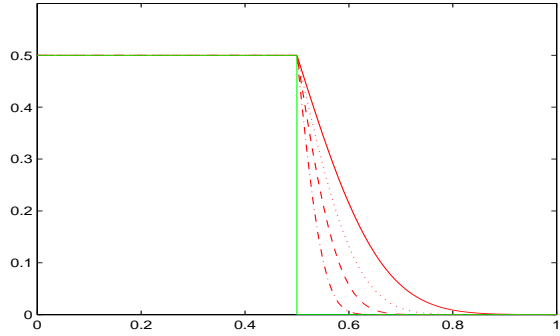}}
\caption{Expected $p$--value curves converging to a stump with increasing $m$.}\label{pvalconstump}
\end{figure}

To introduce and motivate the proposed $p$--value procedure, we first consider the case with $\tau_0$ known. Specifically, let $Y = \mu(X) + \epsilon$, where $\mu$ is a function on $[0,1]$ and
\begin{eqnarray}\label{eq:model}
\mu(x) = \tau_0 \mbox{ for } x \leq d^0, \mbox{ and } \mu(x) > \tau_0  \mbox{ for } x > d^0,
\end{eqnarray}
for $d^0 \in (0,1)$. {\it Note that no other assumptions are made on the behavior of $\mu$ around $d^0$}. The covariate $X$ is sampled from a Lebesgue density $p_X$ on $[0,1]$ and $\epsilon$ is independent of $X$ and distributed as $N(0,\sigma^2)$, with $\sigma$ known. Relaxations of some of these assumptions will be discussed later.
\newline
\newline
As argued above, estimation of $d^0$ via direct estimation of $\mu$ is a hard problem. The use of $p$--values allows a relatively easy solution when multiple responses can be sampled at each covariate value. More specifically, we have:
\begin{eqnarray}
\label{basic-reg-model}
Y_{ij} = \mu(X_{i}) + \epsilon_{ij}, \ \ i = 1,2,\ldots,n;\ \  j = 1,2,\ldots,m,
\end{eqnarray}
with $N = m \times n$ being the total budget of samples. The $\epsilon_{ij}$'s are i.i.d. and distributed like $\epsilon$ above and the $X_i$'s are i.i.d. from $p_X$. The assumption of equal number of replicates ($m$) at each $X_i$ can actually be relaxed to a certain extent and is discussed briefly in the concluding discussion, but for the sake of ease of exposition of the key ideas we assume this throughout the paper.
\newline
\newline
At dose $X_i = x$, we test the null hypothesis $H_{0,x}: \mu(x) = \tau_0$ against the alternative $H_{1,x}: \mu(x) > \tau_0$ using the test statistic $T(x) = \frac{\sqrt{m}(\bar Y_{i,\cdot} - \tau_0)}{\sigma}$ where $\bar Y_{i,\cdot} = \frac{1}{m} \sum_{j=1}^m Y_{ij}$. The observed $p$--value for this test is $p^{(m)}(x) = 1 - \Phi(T(x))$, since $T(x)$ is distributed as $N(0,1)$ under the null hypothesis. From the $n$ different dose levels, we obtain $n$ $p$--values $p^{(m)}(X_1),p^{(m)}(X_2), \ldots,p^{(m)}(X_n)$.
\newline
\newline
Under the null hypothesis which holds to the left of $d^0$, the $p$--values have a Uniform(0,1) distribution. To the right of $d^0$, where the null hypothesis fails, the distributions of the $p$--values change and as $m$ becomes large, the $p$--values converge to the degenerate value $0$. This dichotomous behavior of the $p$--values on either side of $d^0$ can be used to prescribe \emph{consistent} estimates of the latter. A natural way to capture this discrepancy, which we explore in this paper, is to consider the expected $p$--value curve at stage $m$, formally $\nu_m(x) \equiv E(p^{(m)}(x))$. Notice that this is identically 0.5 for all $x \leq d^0$, irrespective of $m$, while for $x > d^0$, it converges to 0 as $m$ increases. We illustrate this in Figure \ref{pvalconstump} assuming $\sigma = 0.5$ for $m =$ 10, 20, 50 and 100. This simple observation can be used to construct estimates of $d^0$ which do not involve estimating $\mu$. We can fit a stump to the observed $p$--values, with levels 1/2 and 0 on either side of the break--point and prescribe the break--point of the best fitting stump (in the sense of least squares) as an estimate of $d^0$. The virtue of this approach lies in the fact that we are able to estimate the threshold consistently, as established rigorously below, by merely fitting a simple mis--specified working model. Its success relies on the fact that the $p$--values eventually show stump like (dichotomous) behavior which no estimate of $\mu$ could have inherited, regardless of sample size. We describe our approach quantitatively below.
\newline
\newline
For convenience, denote $Z_{im} \equiv  p^{(m)}(X_i)$.  Letting $\xi_d(x) = \frac{1}{2} 1(x \le d)$, to find the stump with levels 1/2 and 0 that best approximates the observed $p$--values $Z_{im}, i = 1, 2, \ldots, n$, we  minimize
\begin{eqnarray}\label{eq:CPSSE}
\mathbb{M}_{m,n}(d) = \sum_{i=1}^n \{Z_{im} - \xi_d(X_i)\}^2 =
\sum_{i: X_i \le d} \left( Z_{im} - \frac{1}{2} \right)^2 +  \sum_{i: X_i > d} Z_{im}^2
\end{eqnarray}
over $d \in [0,1]$. Let $\hat d_{m,n} = \arg \min_{d \in [0,1]} \mathbb{M}_{m,n}(d)$, which is a natural estimator of $d^0$.
\newline
\newline
However, in practice, the use of 1/2 and 0 as the stump levels may not always be the best strategy. Firstly, the $p$--values to the right of $d^0$ may not be small enough to be well approximated by 0 for a finite $m$; secondly, we may often have situations where the $Z_{im}$'s are \emph{not exact but approximate} $p$--values. For example, if $\sigma$ was unknown in the above setting,  $T(x)$ would take the form $\sqrt{m}\, \overline{Y}_{i \cdot}/\hat{\sigma}$, where $\hat{\sigma}$ is some estimate of $\sigma$, in which case the $Z_{im}$'s would not be uniformly distributed (or even close to a uniform for modest $m$). In such cases, one can adopt a more adaptive approach by allowing the stump levels to converge to 1/2 and 0 with increasing $m$, or by keeping the stump-levels unspecified and estimating them from the data itself. Such adaptive procedures might provide a better fit to the observed $p$--values and improve the precision of the estimate of $d^0$.
\newline
\newline
We summarize next the proposed estimation procedure and establish its consistency. Formally, the setup is as follows: Consider the (possibly heteroscedastic) regression model $Y = \mu(X) + \epsilon$ with $\mu$ as in (\ref{eq:model}), $E(\epsilon | X) = 0$, $\mbox{Var}(\epsilon | X) > 0$ and the covariate $X$ following a Lebesgue density $p_X$ on $[0,1]$. The available data from this model $\{X_i,\{Y_{ij}\}_{j=1}^m\}_{i=1}^n$ are exactly as in (\ref{basic-reg-model}). The steps of the procedure are the following:
\begin{enumerate}
\item For $i = 1, 2, \ldots, n$, let $p^{(m)}(X_i) \equiv Z_{im}$ denote the observed (potentially approximate) $p$--value based on a test of the hypothesis $H_{0,i}: \mu(X_i) = 0$ against the alternative $H_{1,i}: \mu(X_i) > 0$, using data $\{Y_{ij}: j = 1,2,\ldots,m\}$, such that $Z_{1m}, Z_{2m}, \ldots, Z_{nm}$ are i.i.d. as well.

\item Fit a stump $\alpha_m\,1(x \leq d) + \beta_m\,1(x > d)$ to $\{Z_{im}\}_{i=1}^n$, where $\alpha_m$ and $\beta_m$ are known non-negative quantities that converge to 1/2 and 0 respectively. For $d \in [0,1]$ define:
\begin{eqnarray}
\label{eq:ememendee}
\mathbb{M}_{m,n}(d) = \sum_{i: X_i \le d} (Z_{im} - \alpha_m)^2 +  \sum_{i: X_i > d} (Z_{im} - \beta_m)^2
\end{eqnarray}
and let $\hat d_{m,n} = \arg \min_{d \in [0,1]}\,\mathbb{M}_{m,n}(d)$. We can choose $\alpha_m = 1/2$ and $\beta_m = 0$ for all $m$ to get back to the setting of (\ref{eq:CPSSE}).

\item We can even let the data choose the optimal $\alpha_m$ and $\beta_m$ by setting
    \[ \hat{\theta}_{m,n} \equiv (\hat{\alpha}_{m,n}, \hat{\beta}_{m,n}, \hat{d}_{m,n}) = \arg \min_{\theta \equiv (\alpha,\beta,d) \in [0,1]^3}\;\sum_{i=1}^n\,\{Z_{im} - \alpha\,1(X_i \leq d) - \beta\,1(X_i > d)\}^2 \,.\]
\end{enumerate}

\begin{figure}[!h]
\centering
\resizebox{3.5in}{2in}{\includegraphics{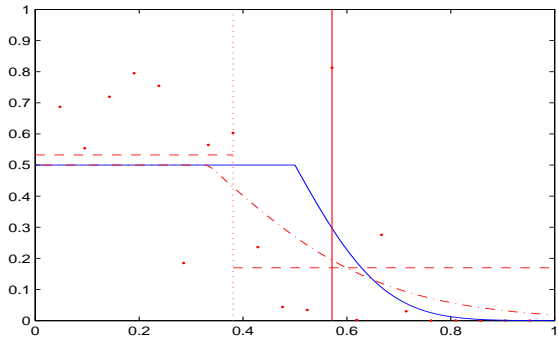}}
\caption{Fitting different models to the $p$--value curve.}\label{fitmodelsdata}
\end{figure}

\begin{theorem}
\label{consistency--theorem}
Consider the above setup of the problem and let $\nu_m(x) = E(Z_{1m} | X_1 = x)$. Assume further that (a)
$\nu_m(x) \rightarrow \nu(x): = (1/2)\,1(x \leq d^0)$ for each $x$, as $m \rightarrow \infty$, and
(b) $p_X(x) > \kappa > 0$ for $x \in [d^0 - l, d^0 + l]$ for some (small) $l > 0$. We then have:
\begin{enumerate}
\item[(i)] $\hat{d}_{m,n} \stackrel{p}{\rightarrow} d^0 \ \ \mbox{as} \ \ m,n \rightarrow \infty,$ i.e., given $\epsilon, \eta > 0$, there exists a positive integer $K$, such that for $m,n \geq K$, $P\{| \hat{d}_{m,n} - d^0 | > \epsilon\} < \eta$.

\item[(ii)] $\hat{\theta}_{m,n} \stackrel{p}{\rightarrow} \theta_0 \equiv (1/2, 0, d^0) \ \ \mbox{as} \ \ m,n \rightarrow \infty.$
\end{enumerate}
\end{theorem}
\noindent
The theorem is proved in the Appendix.
\newline
\newline
{\bf Flexible modeling of the $p$--value curve:} In many applications the curve $\mu$ is an increasing continuous function, whence the expected $p$--value curves are continuous decreasing functions converging to a stump with increasing $m$ as in Figure (\ref{pvalconstump}). One can then also use continuous parametric working models to take into account the shape of the $p$--value curve for finite $m$. Figure (\ref{pvalconstump}) suggests looking at sigmoidal curves. We propose and explore one such family, a 2--parameter sigmoidal family $\mathcal{G}$ next:
\begin{eqnarray}\label{eq:DefPsi}
\mathcal{G} = \left\{ \psi_{d,\alpha}(x) \equiv \frac{1}{2} \mathbf{1}\{x \le d\} + \frac{e^{-\alpha(x-d)}}{1 + e^{-\alpha(x-d)}} \mathbf{1}\{x > d\}: d \in [0,1], \alpha \ge 0 \right\}.
\end{eqnarray}
For $x \le d$, $\psi_{d,\alpha}(x) = 1/2$ and this models the region where $\mu =0$ (or a constant). For $ x>d$, $\psi_{d,\alpha}$ is decreasing, thus mimicking the finite sample behavior of the expected $p$--value curve $\nu_m$. It can be shown that consistent estimates of $d^0$ may be obtained with this misspecified class of models, as well. Consistent estimation of $d^0$ is possible because in the limit (as $\alpha \rightarrow \infty$ and $d \rightarrow d^0$) the sigmoid converges to the stump (indicator) function $\nu$.
\newline
\newline
We estimate the parameters $d$ and $\alpha$ of the working model by solving a least squares problem, i.e.,
\begin{eqnarray}\label{eq:Def_d_alp}
    (\tilde d_{m,n}, \tilde \alpha_{m,n}) = \arg \min_{d \in [0,1], \alpha \ge 0} \mathbb{G}_{m,n}(d,\alpha) \equiv \frac{1}{n} \sum_{i=1}^n \{Z_{im} - \psi_{d,\alpha}(X_i)\}^2.
\end{eqnarray}
It can be shown under assumptions (a) and (b) of Theorem \ref{consistency--theorem} that $\tilde d_{m,n} \stackrel{p}{\rightarrow} d^0$, as $n,m {\rightarrow} \infty$, using arguments similar to those in the proof of that theorem.
\newline
\newline
It should be noted that there is nothing special about using the parametric model (\ref{eq:DefPsi}) to estimate $d^0$. Any reasonable class of models that includes (or can converge to) the stump function would, in principle, yield consistent estimators of $d^0$. However, we focus on the stump $\xi_d$, mostly because of the conceptual/computational simplicity and accurate finite sample performance (as illustrated in the simulation study in Section \ref{sec:simulations}, in which results for the 2--parameter sigmoidal family are also included). Figure \ref{fitmodelsdata} shows the true expected p--value curve (shown by the solid curve) and illustrates the three methods of finding $d^0$ in a single simulation run with $\sigma = 0.5$, $d^0 = 0.5, n = 20, m = 10$: (1) fitting a stump with levels 1/2 and 0 (shown by the solid vertical line denoting the estimated value of $d^0$); (2) fitting a stump with adaptive levels as in Theorem \ref{consistency--theorem} $(ii)$ (shown by the dashed horizontal lines) and the estimated $d^0$ (shown by the dotted vertical line); (3) the fitted sigmoid model defined in (\ref{eq:Def_d_alp}) (shown by the dashed-dotted curve).
\subsection{The procedure under relaxed assumptions}
\label{relaxsec}
While the assumptions of normality of errors and known variance were used to motivate the procedure of stump based approximations to the $p$--values in the previous section, Theorem \ref{consistency--theorem} shows that these assumptions can be relaxed considerably. Further, as the theorem does not require the responses to be continuous, the procedure is also valid in discrete response settings, one of which we illustrate in the subsequent discussion. We discuss below a number of different scenarios that can arise in practice.
\newline
\newline
\emph{(i) Homoscedastic errors with unknown variance:}  Suppose that the $\epsilon_{ij}$'s are continuous i.i.d. random variables with mean 0 and unknown variance $\sigma^2$ and independent of $\{X_i\}_{i=1}^n$. First, define ``working'' $p$--values $p^{(m)}(X_1), p^{(m)}(X_2), \ldots, p^{(m)}(X_n)$ as $p^{(m)}(X_i) = 1 - \Phi\left(\frac{\sqrt{m} (\bar Y_{i,\cdot} - \tau_0)}{\hat{\sigma}(X_i)} \right)$  where $\hat{\sigma}^2(X_i) = \sum_{j=1}^m (Y_{ij} - \overline{Y}_{i \cdot})^2/(m-1)$. The $p$--values are then independent, $\nu_m(x) = E (p^{(m)}(X_1) | X_1 = x )$ converges to $\nu(x)$ and Theorem 2.1 applies, yielding the consistency of the stump-based least squares estimates of $d^0$. However, owing to the homoscedasticity of the errors, the constructed $p$--values are clearly sub-optimal since we could have replaced $\hat{\sigma}(X_i)$ in the definition of the $i$'th $p$--value by $\hat{\sigma}_{m,n}$, where $\hat{\sigma}_{m,n}^2 = \sum_{i=1}^n\,\sum_{j=1}^m\,(Y_{ij} - \overline{Y}_{i,. \cdot})^2/(mn-n)$ is the standard pooled estimate of $\sigma^2$ (and is significantly superior to each $\hat{\sigma}(X_i)$). Thus, least squares estimates of $d^0$ based on $(b)$ and $(c)$ in the build--up to Theorem \ref{consistency--theorem}, using such \emph{improved} $p$--values would certainly yield consistent estimates of $d^0$. Unfortunately, we are no longer quite in the setting of Theorem \ref{consistency--theorem}, as the presence of $\hat{\sigma}$ in each $p$--value makes them dependent, and therefore cannot invoke its conclusions to deduce consistency. We tackle the consistency issue for this problem below. Roughly speaking, as $m,n$ grow, $\hat{\sigma}$ stabilizes quickly, the $p$--values become approximately independent and the scenario approaches that of Theorem \ref{consistency--theorem}.
\begin{theorem}
\label{homosced-err-pvalcons}
Assume that $(a)$ $\inf_{d^0 + \beta \leq d \leq 1} \mu(d) > \tau_0$ for any $\beta>0$, and that $(b)$ $p_X(x) < K$ for $x \in [d^0, d^0 + \eta)$ for some $\eta > 0$. Let $\hat{d}_{m,n}$ be the least squares estimate of $d^0$ obtained as: $\hat{d}_{m,n}= \arg \min_{d}\,\left[\sum_{i: X_i \leq d}\, \left( Z_{im} - \frac{1}{2} \right)^2 + \right.$ $\left. \sum_{i: X_i > d}\, \left( Z_{im} - 0 \right)^2 \,\right]$, where $Z_{im} = 1 - \Phi\left(\frac{\sqrt{m} (\bar Y_{i,\cdot} - \tau_0)}{\hat{\sigma}_{m,n}} \right)$; these being the `improved' $p$--values alluded to in (i) above. Then $\hat{d}_{m,n} \stackrel{p}{\rightarrow} d^0$ as $m,n \rightarrow \infty$.
\end{theorem}
A condensed version of the proof is available in the Appendix.
\newline
\newline
{\bf Remark:} If we \emph{know} that the errors are normal, the statistics $\sqrt{m}\,\overline{Y}_{i \cdot}/\hat{\sigma}$ follow a $t_{mn-n}$ distribution. Letting $F_{t_{mn-n}}$ denote the corresponding distribution function, we can take $Z_{im} = 1 - F_{t_{mn-n}}(\frac{\sqrt{m} \bar Y_{i\cdot}}{\hat{\sigma}})$ and the corresponding estimates of $d^0$ continue to be consistent.
\newline
\newline
\emph{(ii) Heteroscedastic errors:} Consider a regression set-up with continuous responses and heteroscedastic errors: i.e. $\sigma^2(x) \equiv  \mbox{Var}(\epsilon | X=x)$ varies with $x$. Thus, $\sigma^2(X_i)$ is estimated by $\hat{\sigma}^2(X_i) = \sum_{i=1}^m\,(Y_{ij} - \overline{Y}_{i \cdot})^2/(m-1)$ and $Z_{im} = 1 - \Phi \left( \frac{\sqrt{m} (\bar Y_{i,\cdot} - \tau_0)}{\hat{ \sigma}(X_i)} \right)$; in this case, pooling is no longer possible unlike case $(i)$. It is not difficult to see that $\nu_m(x) = E(Z_{im} | X_i = x)$ converges to $\nu(x)$ as above and the conclusions of Theorem \ref{consistency--theorem} hold. In case the errors are conditionally normally distributed, $\Phi$ in the definition of the $p$--values can be replaced by $F_{t_{m-1}}$, the distribution function of the $t_{m-1}$ distribution.
\newline
\newline
\emph{(iii) Discrete responses:} Settings where the response is discrete can also be considered. One can think of the covariate values as dose--levels and suppose, at each dose level, $X_i$, that a binomial experiment with $m$ independent subjects is performed, and the response $Y_i$ is the number that show a reaction to the dose. The function $\mu(x)$ is the probability that a subject yields a reaction at dose $x$ and is assumed to be at a baseline value $p_0 > 0$ for $x \leq d^0$ and greater than $p_0$ otherwise (note that $p_0 = 0$ gives a pathological situation).
\newline
\newline
We base our $p$--value at dose level $X_i$ on the normalized statistic $(Y_i  - m\,p_0)/\sqrt{m\,p_0\,(1 - p_0)}$ with (working $p$--value) $Z_{im} = 1 - \Phi((Y_i  - m\,p_0)/\sqrt{m\,p_0\,(1-p_0)})$. Alternatively, the $Z_{im}$'s could also be defined to be $p$--values based on the exact binomial distribution: i.e., $Z_{im} = 1 - F_{m,p_0}(Y_i)$ where $F_{m,p_0}$ is the distribution function of Binomial$(m,p_0)$. The conditions of Theorem \ref{consistency--theorem} are easy to verify in either case and the conclusions of the theorem continue to hold.

\subsection{A Digression: Composite Hypotheses}\label{comp-hyp}
Consider now a situation where $\mu(x)$ is known to be strictly less than (\emph{a known}) $\zeta_0$ if $x < d^0$ and strictly greater than $\zeta_0$ for $x > d^0$. We keep the behavior at $d^0$ unspecified. For a monotone function $\mu$ this reduces to estimating its inverse at $\zeta_0$.
The latter problem has been well--studied in the literature (Banerjee and Wellner (2005), Tang et. al. (2010) and references therein)
but only under explicit shape and/or smoothness constraints on $\mu$.
Our formulation, on the other hand, is much broader in scope as it requires neither monotonicity, nor smoothness assumptions.
\newline
\newline
To formulate the problem, consider the model posited in (\ref{basic-reg-model}) with homoscedastic normal errors and $\mu$ as in the beginning of the previous paragraph. At dose $X_i = x$ the statistic $T(x) = \sqrt{m}(\overline{Y}_{i \cdot} - \zeta_0)/\sigma$ is used to test the composite hypothesis, $H_{0,x}: \mu(x) < \zeta_0$ versus $H_{1,x}: \mu(x) > \zeta_0$, with rejection for large values of $T(x)$. Construct an approximate $p$--value as $p^{(m)}(x) = 1 - \Phi(T(x))$. Define $\nu_m(x) = E [p^{(m)}(x)]$. It is easy to check that $\nu_m(x)$ converges to 1 as $m \rightarrow \infty$ for $x < d^0$ and to 0 for $x > d^0$. This suggests
$$\hat{d}_{m,n} = \arg \min_d\;\mathbb{M}_{m,n}(d) \equiv \sum_{i: X_i \leq d}\,(Z_{im} - 1)^2 + \sum_{i: X_i > d}\,Z_{im}^2$$
as a natural estimate of $d^0$, with $Z_{im} \equiv  p^{(m)}(X_i), i = 1, 2,\ldots,n$.
\newline
\newline
We present an analogue of Theorem \ref{consistency--theorem} (in the current setting) that generalizes the above strategy. We start by summarizing the proposed estimation procedure.
\begin{enumerate}
\item Identical to Step (a) in the build-up to Theorem \ref{consistency--theorem}, except for $p^{(m)}(X_i) \equiv Z_{im}$ denoting the observed (potentially approximate) $p$--value based on a test of the hypothesis $H_{0,i}: \mu(X_i) < \zeta_0$ against the alternative $H_{1,i}: \mu(X_i) > \zeta_0$.

\item Step (b) remains the same as before but with $\alpha_m$ converging to 1.
\item Step (c) remains unaltered.
\end{enumerate}
\begin{theorem}
\label{consistency--theorem--2}
Let $\nu_m(x) = E(Z_{1m} | X_1 = x)$ and further assume that (a) $\nu_m(x) \rightarrow 1$ for each $x < d^0$ and $\nu_m(x) \rightarrow 0$ for each $x > d^0$, as
$m \rightarrow \infty$, and, (b) $p_X(x) > \kappa > 0$ for $x \in [d^0 - l, d^0 + l]$ for some (small) $l > 0$. Then, we have
\begin{enumerate}
\item   $\hat{d}_{m,n} \stackrel{p}{\rightarrow} d^0 \ \ \mbox{as} \ \ m,n \rightarrow \infty$.

\item $ \hat{\theta}_{m,n} \stackrel{p}{\rightarrow} \theta_0 \equiv (1, 0, d^0) \ \ \mbox{as} \ \ m, n \rightarrow \infty$.
\end{enumerate}
\end{theorem}
\noindent
The proof of the theorem is similar to that of Theorem \ref{consistency--theorem} and is therefore skipped.

\section{The Case of an Unknown $\tau_0$} \label{sec:unknown-tau}
In this section, we address the more realistic situation where $\tau_0$ is unknown. As will become apparent in subsequent developments, a number of complications arise in this setting. In order to focus on the main ideas, we confine ourselves to the setting of continuous responses. Throughout this section, we consider the problem of fitting a one-parameter stump with levels 1/2 and 0 on either side of the jump-location, the main parameter of interest.
\newline
\newline
{\bf Homoscedastic errors:} Consider first, the homoscedastic error setting as in $(i)$ of Section \ref{relaxsec}. The consistency of the least squares estimate of $d^0$ in this model by fitting a stump (with levels $1/2$ and $0$ on either side of the jump) to the observed $p$--values  was established in Theorem \ref{homosced-err-pvalcons}. However, with an unknown $\tau_0$, these $p$--values can not be constructed. A natural alternative is to replace $\tau_0$ with some appropriately consistent estimate. The next theorem shows the consistency of the estimate of $d^0$ obtained by fitting a stump-model to the observed $p$--values when an appropriate estimate of $\tau$ is used. For $\sigma, \tau > 0$, define:
\begin{equation}\label{basic-def-zim}
Z_{im}^{\sigma}(\tau) = 1 - \Phi(\sqrt{m}(\overline{Y}_{i \cdot} - \tau)/\sigma)
\end{equation}
for $i = 1,2,\ldots,n$.
\begin{theorem}\label{thm:consTauUnk}
Consider model (\ref{basic-reg-model}) where the errors $\{\epsilon_{ij}\}$ are i.i.d with variance $\sigma_0^2$ and independent of $\{X_i\}$. Suppose that $\tilde{\tau} \equiv \tilde{\tau}_{m,n}$ is a consistent estimator of $\tau_0$ such that $\sqrt{m}(\tilde \tau - \tau_0) = o_p(1)$. Let $\tilde \sigma \equiv \tilde  \sigma_{m,n}$ be a consistent estimator of $\sigma_0$ and let $\tilde{d} \equiv \tilde{d}_{m,n}$ be estimated as:
\begin{equation}
\label{fit-stump-appx-pval}
\tilde{d}_{m,n} = \arg \min_{d \in [0,1]}\,\left[\sum_{i: X_i \leq d}\,\left\{Z_{im}^{\tilde{\sigma}_{m,n}}(\tilde{\tau}_{m,n}) - \frac{1}{2} \right\}^2 + \sum_{i: X_i > d}\,\left\{Z_{im}^{\tilde{\sigma}_{m,n}}(\tilde{\tau}_{m,n}) - 0 \right\}^2 \, \right].
\end{equation}
Then $\tilde{d}_{m,n} \stackrel{p}{\rightarrow} d^0$ as $m,n \rightarrow \infty$.
\end{theorem}
{\bf Remark:} In a situation where $d^0$ may be safely assumed to be greater than some known positive $\eta$, an estimate of $\tau_0$ satisfying the condition of the above theorem can be obtained by taking the average of the response values on the interval $[0,\eta]$. However, this does not offer a satisfactory solution, since such an $\eta$ may not be known in various applications. Also, even if $\eta$ is known but small, the estimate thus obtained will be unsatisfactory unless $n$ is really large. Below, we adopt a more principled approach to the estimation of $\tau_0$ that does not require such background knowledge, once again using $p$--values.
\newline
\newline
We now focus on constructing an explicit estimator $\tilde \tau$ of $\tau_0$ as required in Theorem \ref{thm:consTauUnk}, using $p$--values. Recall the definition of $Z_{im}^{\sigma}(\tau)$ in (\ref{basic-def-zim}). Let $\tau > \tau_0$ and note that as $m$ increases, for $\mu(X_i) < \tau$, $Z_{im}^{\hat{\sigma}_{m,n}}(\tau)$ goes to 1 in probability, while for $\mu(X_i) > \tau$, $Z_{im}^{\hat{\sigma}_{m,n}}(\tau)$ goes to 0 in probability. For any $\tau < \tau_0$, it is easy to see that  $Z_{im}^{\hat{\sigma}_{m,n}}(\tau)$ always goes to 0 in probability, whereas when $\tau = \tau_0$, $Z_{im}^{\hat{\sigma}_{m,n}}(\tau)$ goes to 0 for $X_i > d^0$, but is uniformly distributed on $(0,1)$ for $X_i < d^0$ for every $m$. Thus, it is only when $\tau = \tau_0$ that $Z_{im}^{\hat{\sigma}_{m,n}}(\tau)$'s are the closest to $1/2$ for a substantial number of $i$'s. This suggests a natural estimate for $\tau_0$: namely,
\begin{equation}\label{eq:Tau_mn}
\hat{\tau}_{m,n} = \arg \min_\tau \sum_{i=1}^n \{Z_{im}^{\hat{ \sigma}_{m,n}} (\tau)- 1/2\}^2.
\end{equation}
Once $\hat{\tau}_{m,n}$ is obtained, an estimate of $d^0$, say $\hat{d}_{m,n}$ can be obtained by taking $\tilde{\tau}_{m,n}$ to be $\hat{\tau}_{m,n}$ in (\ref{fit-stump-appx-pval}) and $\tilde{\sigma}_{m,n}$ to be $\hat{\sigma}_{m,n}$. This method of estimating $d^0$ and $\tau_0$ is referred to, subsequently, as {\it Method 1}. Theorem \ref{tau-cons-theorem} shows that under some mild conditions on the function $\mu$, $\sqrt{m}\,(\hat{\tau}_{m,n} - \tau_0)$ is $o_p(1)$. This along with the fact that $\hat{\sigma}_{m,n}$ is consistent for $\sigma_0$, implies that $\tilde d_{m,n}$ is consistent for $d^0$ by Theorem \ref{thm:consTauUnk}.
\begin{theorem}
\label{tau-cons-theorem}
Consider the same setup as in Theorem \ref{thm:consTauUnk}. Further suppose that the regression function $\mu$ satisfies:
\begin{itemize}
\item[(A)] Given $\eta > 0$, there exists $\epsilon > 0$ such that, for every $\tau > \tau_0$, $$\int_{\{x > d^0 : |\mu(x) - \tau| \le \epsilon\}} p_X(x) dx < \eta.$$
\end{itemize}
Also assume that $\phi_m$, the density function of $\sqrt{m}\,\overline{\epsilon}_{1.}/\sigma_0$, converges pointwise to $\phi$, the standard normal density.
Then $\sqrt{m}\,(\hat{\tau}_{m,n} - \tau_0) \stackrel{p}{\rightarrow} 0$ as $m,n \rightarrow \infty$.
\end{theorem}
{\bf Remark:} Condition (A) is guaranteed if, for example, $\mu$ is strictly increasing to the right of $d^0$ although they hold under weaker assumptions on $\mu$. In particular it rules out flat stretches to the right of $d^0$. Note that the assumption that $\phi_m$ converges to $\phi$ is not artificial, since convergence of the corresponding distribution functions to the cdf of the standard normal is guaranteed by the central limit theorem.
\newline
\newline
{An alternative method ({\it Method 2}):} Notice that the previous method involves the estimation of $d^0$ in two steps: first by estimating $\tau_0$ and subsequently using this estimate to approximate the $p$--values to which a stump is fitted, as described in (\ref{fit-stump-appx-pval}). An alternative {\em one-step} method for estimating $d^0$ (that avoids estimating $\tau_0$) is presented, when \emph{$\mu$ is increasing}. Define $\xi_n(x) = E(Y | X \leq x)$. A natural estimate of $\xi_n(x)$ is given by $\hat{\xi}_n(x) = \sum_{i: X_i \leq x} \sum_{j=1}^m\,Y_{ij}/\{m\,\sum\,1(X_i \leq x)\}$. It can be easily checked that for $x \in (0,1)$, $\hat{\xi}_n(x) - \xi(x)$ is $O_p((mn)^{-1/2})$, a fact that will be used later. As our estimate of $d^0$ we propose:
\begin{equation}
\label{alter-method}
\tilde{d}_{m,n} = \arg \min_{d \in [0,1]}\,\left[\sum_{i: X_i \leq d}\,\{Z_{im}^{\hat{\sigma}_{m,n}}(\hat{\xi}_n({X_i})) - 1/2\}^2 + \sum_{i: X_i > d}\,\{Z_{im}^{\hat{\sigma}_{m,n}}(\hat{\xi}_n(X_i)) - 0\}^2 \, \right].
\end{equation}
We do not formally establish the consistency of this procedure in the Appendix, but provide a heuristic discussion below. The performance of this method is also assessed via simulation studies. Once $\tilde{d}_{m,n}$ has been obtained, an estimate of $\tau_0$ is given by $\hat{\xi}_n(\tilde{d}_{m,n})$.
\newline
\newline
{\bf Discussion:} Denote the quantity within the big square brackets on the right side of the above display by $\Psi(d)$. Let $\eta > 0$ be such that $0 < d^0 - \eta < d^0 + \eta < 1$. Consider the difference $\Psi(d^0 - \eta) - \Psi(d^0)$, which can be written as: \[ \sum_{d^0 - \eta < X_i \leq d^0}\, \left[\{Z_{im}^{\hat{\sigma}_{m,n}}(\hat{\xi}_n(X_i))\}^2 - \{Z_{im}^{\hat{\sigma}_{m,n}} (\hat{\xi}_n(X_i)) - 0.5\}^2 \right] \,.\]
Now, for any $X_i \in (d^0 - \eta, d^0]$,
\[ Z_{im}^{\hat{\sigma}_{m,n}}(\hat{\xi}_n(X_i)) = 1 - \Phi\,\left(\frac{\sqrt{m}(\overline{Y}_{im} - \tau_0)}{\hat{\sigma}_{m,n}}  + \frac{\sqrt{m}\, (\tau_0 - \hat{\xi}_n(X_i))}{\hat{\sigma}_{m,n}} \right) \,.\]
For sufficiently large $m,n$, the first term within the brackets on the right side of the above display is approximately distributed like a standard normal, while the second term is small by virtue of the fact that $\sqrt{m}\,(\hat{\xi}_n(X_i) - \tau_0)$ is $O_p(n^{-1/2})$ (where we tacitly make use of the fact that these $X_i$'s are all bounded away from 0). It follows that the right side is approximately distributed like a Uniform(0,1) denoted by $U_i$. Thus, $\Psi(d^0 - \eta) - \Psi(d^0)$ behaves approximately like $\sum_{i: d^0 - \eta < X_i \leq d^0}\, U_i^2 - \sum_{i: d^0 - \eta < X_i \leq d^0}\, (U_i - 0.5)^2$ for $U_i$'s that are approximately uniform and weakly correlated for sufficiently large $m,n$. But this quantity will tend to be non-negative with high probability. A similar argument can be used to show that $\Psi(d^0 + \eta) - \Psi(d^0)$ will tend to be non-negative with high probability, when $\mu$ is increasing, which we leave to the reader. This illustrates why the minimizer of $\Psi$ is close to $d^0$, with high probability, in the long run.
\newline
\newline
{\bf Heteroscedastic Errors:} We briefly discuss the case of heteroscedastic errors as in $(ii)$ of Subsection \ref{relaxsec}. Let $\hat{\sigma}_m^2(X_i) \equiv \sum_{j=1}^m\,(Y_{ij} - \overline{Y}_{i \cdot})^2/(m - 1)$ and $Z_{im}^{\hat{\sigma}_m(X_i)}(\tau) = 1 - \Phi(\sqrt{m}\, (\overline{Y}_{i \cdot} - \tau)/\hat{\sigma}_m(X_i))$. To construct a consistent estimator $\check \tau$ of $\tau_0$ we can use (\ref{eq:Tau_mn}) with $Z_{im}^{\hat \sigma_{m,n}}(\tau)$ changed to $Z_{im}^{\hat \sigma_{m}(X_i)}(\tau)$. Now, Method 1 can be implemented to obtain an estimator for $d^0$ using (\ref{fit-stump-appx-pval}) with $Z_{im}^{\hat \sigma_{m,n}}(\tilde \tau_{m,n})$ replaced by $Z_{im}^{\hat \sigma_{m}(X_i)}(\check \tau)$. Method 2 can also be implemented by replacing the superscript $\hat \sigma_{m,n}$ by $\hat{\sigma}_m(X_i)$ in (\ref{alter-method}).

\section{Multiple Change Points}\label{multiple-pts}
While our procedure was primarily motivated by applications with single baseline thresholds, it can be extended without much difficulty to the case of multiple thresholds. To illustrate the idea, consider a regression model where the function $\mu(x)$, with $x$ varying in $(0,1)$ is at its baseline value, say $\tau_0$, on an interval of the form $[a,b]$ with $0 < a < b <1$ and stays above the baseline elsewhere.
For ease of illustration, we restrict ourselves to the situation with a continuous response and homoscedastic errors with unknown variance, as in Section 2.1 (i). As in that problem, we would construct $p$-values at each point, $\{Z_{im}\}_{i=1}^n$. Our estimates of $a$ and $b$ would be obtained as:
\[ (\hat{a}_{m,n}, \hat{b}_{m,n}) = \arg \min_{a < b}\,\left[ \sum_{X_i \in [a,b]}\,(Z_{im} - 1/2)^2 + \sum_{X_i \notin [a,b]}\,Z_{im}^2 \right] \,.\] The computation of the minimizer would proceed by searching over all pairs $(X_{(i)}, X_{(j)})$ with $i < j$. By extending the techniques used in our proofs
one can establish consistency of the above least squares estimates without much additional difficulty. In case $\tau_0$ is unknown, one can use exactly the same estimate of $\tau_0$ as advocated in (\ref{eq:Tau_mn}). Similar reasoning as in the discussion preceding (\ref{eq:Tau_mn}) shows that $\hat{\tau}_{m,n}$ would be consistent for $\tau_0$ in this setting. It is not difficult to see how the procedure would extend to other kinds of regions, for example, a baseline zone which consists of a disjoint union of finitely many intervals, though the computational complexity increases rapidly with the number of disjoint intervals concerned. The structure of the baseline zone will, of course, be determined by the application under consideration.
\newline
\newline
The procedure can also be extended to allow for situations where $\mu$ attains its minimum and maximum values, say $\tau_{min}$ and $\tau_{max}$, on disjoint intervals $[a,b]$ and $[c,d]$, as is the case for the gene expression data. Assume, for simplicity, that the minimum and maximum values are known (or that very reliable estimates are available). One then applies the procedure in the previous paragraph to determine $[a,b]$. To determine $[c,d]$, note that in the model $-Y_i = -\mu(x) - \epsilon_i$, $[c,d]$ is the region on which $-\mu$ hits its minimum and therefore the above procedure can once again be applied with the signs of the responses flipped.
\newline
\newline
When the minimum and the maximum are unknown,
a natural temptation would be to estimate the minimum via (\ref{eq:Tau_mn}) in the original problem, and the maximum via  (\ref{eq:Tau_mn}) again in the sign-flipped problem, separately; however, that becomes suspect in this situation since $\mu$ has two flat stretches while the method advocated in (9) is theoretically justified only when flat stretches at levels larger than the minimum are ruled out; see Assumption (A) of Theorem \ref{tau-cons-theorem} in this context. However, in many situations, there will be a reasonable degree of separation between the minimum and the maximum and it will be possible to identify these values up to mutually disjoint intervals (for more on this issue see the analysis of the gene data example). In such cases, the procedure in (9) can again be brought into play by doing \emph{restricted searches} over the $\tau$ domain: to identify the minimum, minimize the criterion function in (\ref{eq:Tau_mn}) \emph{only over the interval in which the minimum is expected to lie} (as opposed to searching over the entire putative range of $\mu$) and do a similar analysis for the maximum (by switching to the sign-flipped problem). In the absence of other flat stretches (besides the minimum and the maximum) this procedure will estimate the extremal values accurately.

\section{Simulation Results and Data Analysis}\label{sec:simulations}
We first study the performance of our proposed methods through an extensive simulation study. We then compare our approach with other competing procedures mostly developed in the dose--response setting. The methodology is subsequently illustrated on the motivating examples of the complex queueing system and the gene expression data. In the simulation study, we compare in a number of settings, the simplest model of the one-parameter stump to both the more complex, sigmoid and the three--parameter adaptive stump models. We undertake a comprehensive evaluation of the two methods proposed for estimating the threshold $d^0$ in the presence of unknown $\tau_0$, we investigate the performance of the proposed methodology when the threshold is located close to the boundary of the design space and finally we discuss how one should allocate a fixed budget of samples between number of doses $n$ and replicates $m$.

\subsection{Simulation Studies}
\begin{figure}[!]
\centering
\resizebox{5in}{2in}{\includegraphics{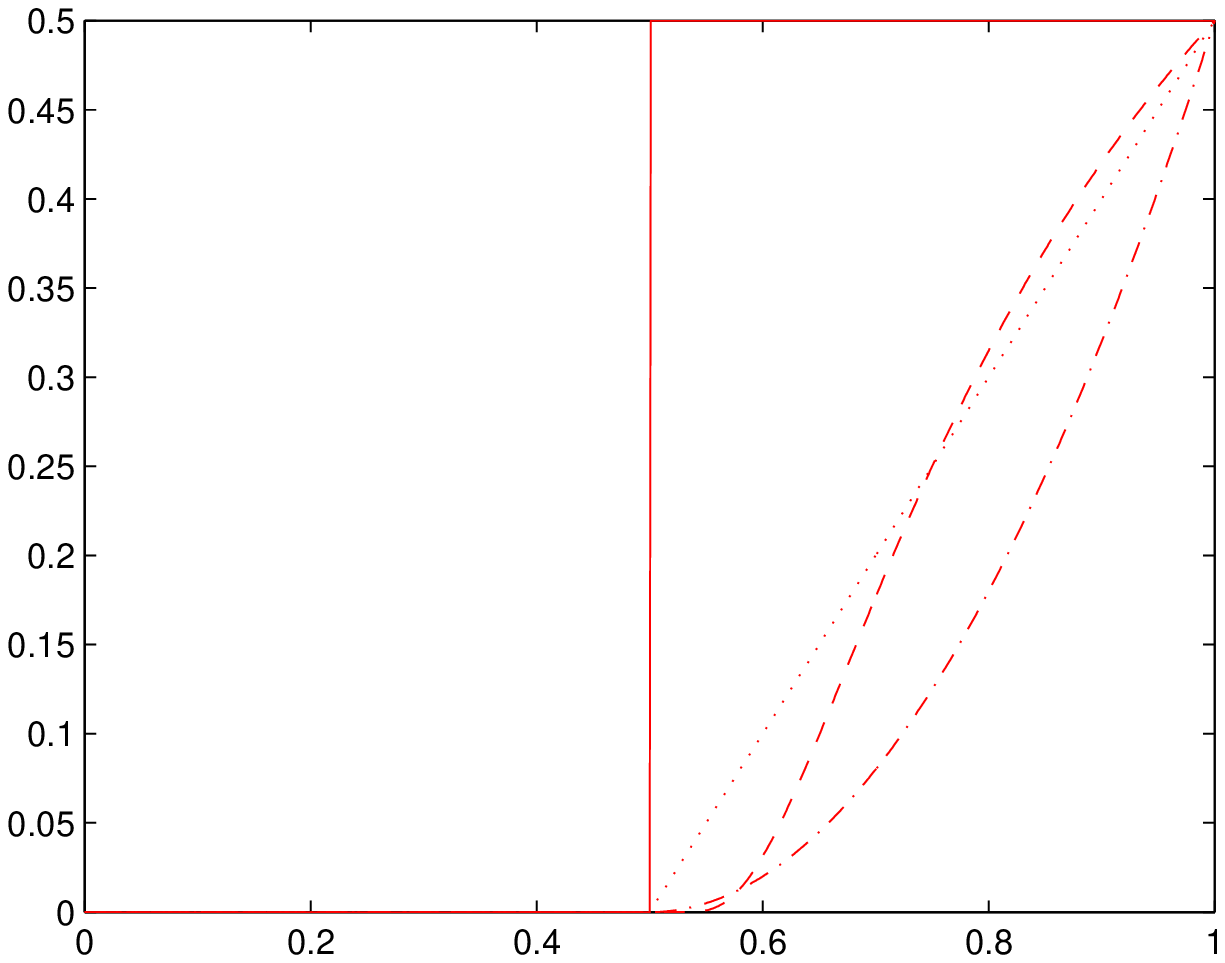} \ \
\includegraphics{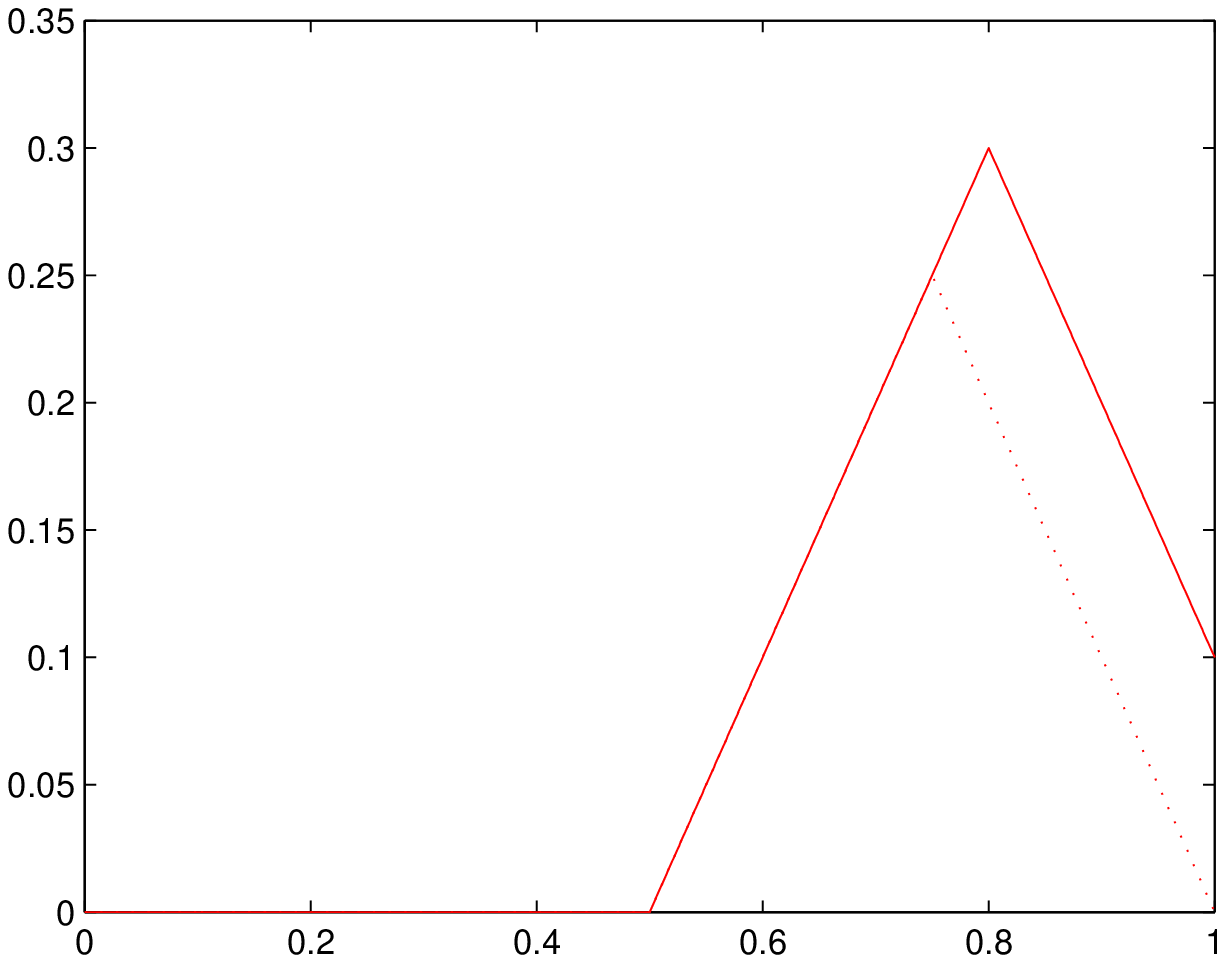}}
\caption{Plots of the four increasing regression functions $\mu$'s (left panel) and the two tent shaped $\mu$'s (right panel).}\label{six-reg-funcs}
\end{figure}

In our numerical studies, six choices of the regression function $\mu$ are considered. Four of these are non-decreasing and the remaining two are ``tent"-shaped. The four monotone ones are depicted in the left-panel of Figure \ref{six-reg-funcs}, all of which are 0 to the left of $d^0 = 0.5$. Specifically, $M_0$, shown by the solid line, is a stump and is identically equal to 0.5 to the right of $d^0$; $M_1$ described by the dotted line is a piece-wise linear function (a kink-model) rising from 0 to 0.5 between $d^0$ and 1; $M_2$, the convex curve, grows like a quadratic beyond $d^0$, whilst $M_3(x) = \exp(-\lambda/(x - 0.5))\,1(x > 0.5)$ (for an appropriate $\lambda$ so that $M_3(1) = 0.5$) is infinitely differentiable at $d^0$. Thus, from $M_0$ to $M_3$ we have four functions exhibiting increasing smoothness at $d^0$.
\newline
\newline
Our two ``tent"--shaped regression functions are depicted in the right panel of Figure \ref{six-reg-funcs}; they are identically 0 till $d^0 = 0.5$ and tent-shaped beyond: $M_4$ rises linearly from $0.5$ to $0.75$ with unit slope and then declines symmetrically from 0.75 to 1, reaching 0 at the point 1, while $M_5$ is a slight variant of $M_4$ rising linearly with unit slope from 0.5 till 0.8 and then decreasing with unit slope from 0.8 till 1.0. Estimation of $d^0 \equiv 0.5$ for the tent-shaped functions is expected to be more challenging than for the monotone ones. Note that for the monotone curves, the ``signal'' keeps on increasing as we move to the right of $d^0$, whereas with the tent-shaped curves, the signal starts to weaken beyond a point, and with $M_4$ especially, comes back to the baseline value at the right boundary.
\newline
\newline
For each allocation pair $(m,n)$ and regression function $\mu(x)$, we generate responses $\{Y_{i1}, Y_{i2},$ $ \ldots, Y_{im}\}$ at covariate value $x_{i} = i/(n+1)$, for $i = 1,2,\ldots,n$, with $Y_{ij} = \mu(x_{i}) + \epsilon_{ij}$, the $\{\epsilon_{ij}\}$'s being i.i.d. $N(0,\sigma^2)$. Two different choices of $\sigma$ (0.1 and 0.3) are considered. For each combination of model, allocation and noise-level $(\sigma)$, the performance, in terms of Root Mean Square Error (henceforth RMSE), of the estimator under consideration is evaluated based on 2000 replicates.  Note that a non-random uniform design is used to generate the covariates, which is a slightly different procedure from generating the $n$ covariates uniformly from $(0,1)$. Of course, asymptotically, it does not make a difference, but for small $n$, the regularity of the uniform grid has a salutary effect on the performance of the least squares estimates.
\newline
\newline
\begin{table}
\caption{RMSEs for the one-parameter stump (first entry of each column) and the two-parameter sigmoid (second entry of each column) with $\sigma = 0.1$ (top) and 0.3 (bottom panel) for five models and different choices of $m$ and $n$.}
\begin{tabular}{|l||c|c|c|c|c|c|}
\hline
$\sigma=0.1$ & &  & & & \\ \hline
$(m,n)$ & $M_0$ &  $M_1$ & $M_2$ & $M_3$ & $M_4$\\
\hline \hline
(5, 5) & 0.153, 0.193 & 0.154,  0.195 & 0.163,  0.217 & 0.154,  0.199 & 0.154,  0.195 \\
(5, 10) & 0.130,  0.170 & 0.130, 0.169 & 0.119, 0.164 & 0.099,  0.150 & 0.116,  0.174 \\
(10, 10) & 0.127, 0.169 & 0.131,  0.171 & 0.110,  0.159 & 0.093,  0.142 & 0.124,  0.170 \\
(10, 20) & 0.079,  0.126 & 0.068,  0.120 & 0.097,  0.126 & 0.075,  0.110 & 0.060,  0.120 \\
(10, 50) & 0.030,  0.066 & 0.033, 0.069 & 0.106,  0.100 & 0.082,  0.082 & 0.033,  0.073 \\
(20, 50) & 0.028, 0.062 & 0.030,  0.066 & 0.088,  0.088 & 0.073,  0.077 & 0.030,  0.071 \\
(50, 100) & 0.014,  0.033 & 0.014,  0.032 & 0.073, 0.058 & 0.071,  0.061 & 0.014,  0.033 \\
\hline
\hline
$\sigma=0.3$ & &  & & & \\ \hline
$(m,n)$ & $M_0$ &  $M_1$ & $M_2$ & $M_3$ & $M_4$\\
\hline \hline
(5, 5)  & 0.154,  0.195 & 0.155, 0.202 & 0.201,  0.235 & 0.171,  0.219 & 0.208,  0.255 \\
(5, 10)  & 0.130,  0.169 & 0.132,  0.175 & 0.200,  0.222 & 0.140,  0.178 & 0.214,  0.258 \\
(10, 10)  & 0.130,  0.164 &0.137,  0.175 & 0.166,  0.198  &0.119,  0.163 & 0.148,  0.226 \\
(10, 20)  & 0.069,  0.115 & 0.089,  0.134 & 0.177,  0.181 & 0.116,  0.135 & 0.120,  0.220 \\
(10, 50)  & 0.032,  0.062 & 0.086,  0.093 & 0.193,  0.153 & 0.128,  0.114 & 0.095,  0.212 \\
(20, 50)  & 0.032, 0.062 & 0.061,  0.079 & 0.162,  0.134 & 0.117,  0.092 & 0.061,  0.135 \\
(50, 100)  & 0.014,  0.033 & 0.039,  0.041 & 0.131,  0.093 & 0.098,  0.081 & 0.039,  0.050 \\
\hline
\end{tabular}
\end{table}
{\bf One-parameter stump vs sigmoid model:} Table 1 gives the RMSEs for estimating $d^0$, for a number of different $(m,n)$ combinations, models $M_0$ through $M_4$ and two values for $\sigma$ (0.1 and 0.3), using a one-parameter stump $(1/2)\,1(x \leq d)$ and a two-parameter sigmoid, as described in (\ref{eq:DefPsi}). The first key observation is that in the relatively high ``signal-to-noise" regime ($\sigma = 0.1$), the inference problem is relatively easier and the one-parameter stump outperforms the two-parameter sigmoid almost uniformly (with an occasional reversal in high $m,n$ settings).  Secondly, $M_2$ and $M_3$ show greater RMSEs in general than $M_0$ and $M_1$ (which could be ascribed to the former two models exhibiting greater smoothness at $d^0$), and interestingly enough greater RMSEs than $M_4$ as well (which is a misspecified model as it returns to 0 at the right end of its support). As expected, increasing both $m$ and $n$ leads to improved performance, by and large. The use of the sigmoidal approximation in the more modest ``signal-to-noise" regime ($\sigma = 0.3$) for combinations of larger $m$ and $n$ leads to better results for the smoother models $M_2$ and $M_3$, but not for $M_4$, since the shape of the sigmoidal curve conflicts badly with that of the corresponding expected $p$--value curve. Further, the RMSEs for $M_2$ generally tend to be worse than $M_3$ for both $\sigma = 0.1$ and $\sigma = 0.3$, even though $M_3$ is smoother in the vicinity of $d^0$. This can be explained by the fact that for the most part to the right of $d^0$, $M_3$ provides more signal than $M_2$, which makes detection easier in the former case. As $m$ and $n$ increase, $M_2$ improves and its performance comes closer to that of $M_3$ (see (50,100) setting).
\newline
\newline
{\bf Comparisons between the one-parameter stump and the three-parameter (adaptive) stump:} We compared the 1--parameter stump (with 0.5 and 0 as the levels on either side of the unknown split-point) to the three-parameter stump, where both levels as well as the split-point are kept unspecified. The comparison is done for three of the six models at two different $\sigma$'s, $0.1$ and $0.3$, for a set of different $(m,n)$ allocations. Although the results are not shown due to space considerations, the results to a large extent are not radically different, and neither method systematically outperforms the other. For $\sigma = 0.1$ and relatively small $m,n$, the adaptive stump exhibits slightly better performance for $M_1$ and $M_5$ and is more or less comparable to the 1-parameter stump for $M_2$. At $\sigma = 0.3$, the advantage of the adaptive stump lessens. Finally, for large $(m,n)$, both types of stumps behave similarly. Hence, in all subsequent simulations one-parameter stumps are employed.
\newline
\newline
\noindent {\bf Assessment of Methods 1 and 2 for estimating $d^0$ when $\tau_0$ is unknown:} We next present a comparison of Methods 1 and 2 from Section \ref{sec:unknown-tau} for estimating $d^0$, when $\tau_0$ is assumed unknown, for two models (due to space considerations): the monotone $M_3$, and the tent-shaped $M_4$. For each model, we present the RMSEs for $d^0$ for each of the methods and also those for $\tau_0$ in Table 2.
\newline
\newline
It is noted that for fixed $\sigma$ and $(m,n)$, the RMSEs for $d^0$ increase from $M_3$ to $M_4$. This phenomenon is also observed when we compare any one of the monotone models $M_1$, $M_2$ or $M_3$ with either of the tent-shaped ones $M_4$ or $M_5$. Since the stump is a monotone working model, this result is expected in light of the first three models and the lack of monotonicity of the others. Overall, Methods 1 and 2 are comparable, although for smaller $\sigma=0.1$, Method 1 exhibits smaller RMSEs for $d^0$, while for larger $\sigma=0.3$, Method 2 dominates slightly. It is worth noting that for model $M_4$, for small $m$, the performance deteriorates with increasing $n$, for large $\sigma$. Although counter--intuitive, this phenomenon can be explained by noticing that in this situation the expected $p$--value curve has a ``V''--shape to the right of $d^0$ and conforms badly with the monotone nature of the fitted stump. For small $n$, this discrepancy is somewhat masked by the small number of observations (i.e., $p$--values), but for large $n$, this non--conformity is clearly exhibited. Regarding $\tau_0$, Method 2 has a slight edge over Method 1 at both noise levels. Finally, for the challenging setting of $M_4$, the performance of both methods is rather inferior for small values of $m$ and $\sigma = 0.3$.
\newline
\newline
\begin{table}
\caption{RMSEs for $d^0$ and $\tau_0$ with $\sigma = 0.1$ and $0.3$, respectively for models $M_3$ and $M_4$ and different choices of $m$ and $n$, using the two proposed methods.}
\label{tunkM3}
\begin{small}
\begin{minipage}[b]{0.5\linewidth}
\centering
\begin{tabular}{|l||c|c|c|c|c|}
\hline
$M_3$ & \multicolumn{4}{c|}{$\sigma=0.1$} \\ \hline
$(m,n)$ & $d^0 (1)$ & $d^0 (2)$ & $\tau_0 (1)$ & $\tau_0 (2)$\\
\hline \hline
(5, 5)  & 0.091 & 0.105 & 0.040 & 0.032 \\
(5, 10)  & 0.086 & 0.094 & 0.027 & 0.023 \\
(10, 10)  & 0.070 & 0.081 & 0.019 & 0.016 \\
(10, 20)  & 0.078 & 0.079 & 0.014 & 0.011 \\
(10, 50)  & 0.087 & 0.084 & 0.009 & 0.006 \\
(20, 50)  & 0.077 & 0.074 & 0.006 & 0.004 \\
(50, 100)  & 0.072 & 0.071 & 0.003 & 0.002 \\
\hline
\end{tabular}
\end{minipage}
\hspace{0.5cm}
\begin{minipage}[b]{0.5\linewidth}
\centering
\begin{tabular}{|l||c|c|c|c|c|}
\hline
$M_3$ & \multicolumn{4}{c|}{$\sigma=0.3$} \\ \hline
$(m,n)$ & $d^0 (1)$ & $d^0 (2)$ & $\tau_0 (1)$ & $\tau_0 (2)$\\
\hline \hline
(5, 5)  & 0.180 & 0.161 & 0.120  & 0.097 \\
(5, 10)  & 0.191 & 0.158 & 0.099 & 0.067 \\
(10, 10)  & 0.137 & 0.121 & 0.067 & 0.048 \\
(10, 20)  & 0.141 & 0.122 & 0.049 & 0.030 \\
(10, 50)  & 0.147 & 0.135 & 0.033 & 0.019 \\
(20, 50)  & 0.120 & 0.112 & 0.021 & 0.013 \\
(50, 100)  & 0.102 & 0.099 & 0.009 & 0.006 \\
\hline
\end{tabular}
\end{minipage}
\end{small}
\end{table}

\begin{table}
\begin{small}
\begin{minipage}[b]{0.5\linewidth}
\centering
\begin{tabular}{|l||c|c|c|c|c|}
\hline
$M_4$ & \multicolumn{4}{c|}{$\sigma=0.1$} \\ \hline
$(m,n)$ & $d^0 (1)$ & $d^0 (2)$ & $\tau_0 (1)$ & $\tau_0 (2)$\\
\hline \hline
(5, 5)  &  0.138 & 0.112 & 0.067 & 0.032 \\
(5, 10)  &  0.109 & 0.113 & 0.045 & 0.026 \\
(10, 10)   & 0.074 & 0.102 & 0.027 & 0.018 \\
(10, 20)  & 0.046 & 0.060 & 0.018 & 0.011 \\
(10, 50)  & 0.034 & 0.033 & 0.012 & 0.007 \\
(20, 50)  & 0.025 & 0.028 & 0.008 & 0.005 \\
(50, 100)  & 0.014 & 0.015 & 0.003 & 0.002 \\
\hline
\end{tabular}
\end{minipage}
\hspace{0.5cm}
\begin{minipage}[b]{0.5\linewidth}
\centering
\begin{tabular}{|l||c|c|c|c|c|}
\hline
$M_4$ & \multicolumn{4}{c|}{$\sigma=0.3$} \\ \hline
$(m,n)$ & $d^0 (1)$ & $d^0 (2)$ & $\tau_0 (1)$ & $\tau_0 (2)$\\
\hline \hline
(5, 5)  &  0.243 & 0.221 & 0.110 & 0.096 \\
(5, 10)  &0.311 & 0.294 & 0.091 & 0.077 \\
(10, 10)  & 0.262 & 0.232 & 0.081 & 0.058 \\
(10, 20)  & 0.286 & 0.270 & 0.064 & 0.047 \\
(10, 50)  & 0.313 & 0.317 & 0.053 & 0.044 \\
(20, 50)  & 0.133 & 0.137 & 0.040 & 0.021 \\
(50, 100)  & 0.050 & 0.040 & 0.016 & 0.006 \\
\hline
\end{tabular}
\end{minipage}
\end{small}
\end{table}
\noindent {\bf Comparisons among the methods for extreme values of $d^0$:} We have so far concentrated on the case with $d^0=0.5$. We investigate next, settings where $d^0$ is closer to one of the boundaries; specifically, where the ``action'' starts fairly quickly at a low covariate level and also where the ``action'' starts late. To this end, we consider the models $\tilde{M}_1$ and $\tilde{M}_2$, where $\tilde{M}_1$ is flat at zero till 0.2 and then rises linearly with slope 1 all the way up to 1, while $\tilde{M}_2$ is flat all the way till 0.8 and then rises linearly with unit slope. These are simple variants of the kink-model $M_1$. We report the performances of the 1 parameter stump when $\tau_0$ is known (to be 0) with (the homoscedastic error variance) $\sigma=0.1$ and $0.3$ for both models, and also the performances of Methods 1 and 2 in the $\tau_0$ unknown case for different $(m,n)$ allocations in Table 3. The design-points are chosen similarly to the case $d^0 = 0.5$.
\newline
\newline
For $\tau_0$ known, the one-parameter stump behaves qualitatively as one would expect. The RMSEs tend to be bigger for the $d^0 = 0.8$ case (for a fixed allocation and noise level), though with increasing $(m,n)$ the RMSEs for both models tend to converge and are similar to the numbers for $M_1$ in the $\tau_0$ known case, since the models $M_1, \tilde{M}_1$ and $\tilde{M}_2$ look exactly similar in a small neighborhood of the kink, and it is this local behavior that drives the asymptotic MSE as $m,n$ go to infinity. More interesting is the case when $\tau_0$ is unknown. In this case, with the model $\tilde{M}_2$, Method 1 generally produces slightly smaller RMSEs for $d^0$ for both $\sigma= 0.1$ and $0.3$ apart from some of the ``rich allocation'' scenarios where the performances of both methods are very comparable. Method 2, on the other hand, gives better RMSEs for $\tau_0$. With Model $\tilde{M}_1$, we see a markedly different phenomenon. For small $(m,n)$ in the case $\sigma = 0.1$ and $\sigma = 0.3$ (and also for some small $m$, large $n$ scenarios in the latter case), Method 1 shows very poor performance compared to Method 2 with much larger RMSEs for $d^0$. A glance at the RMSEs for $\tau_0$ reveals what is happening. The estimates of $\tau_0$ (needed in Method 1 to compute surrogate $p$--values) are extremely poor, and this leads to biased estimates for $d^0$. The poor performance is, of course, exacerbated for higher values of $\sigma$. These results indicate that Method 1 can perform pretty badly for small $n$; in that case, the number of covariate values in the flat stretch is small for $\tilde M_1$ and this affects the estimation of $\tau_0$ badly.
\newline
\newline
\begin{table}
\caption{RMSEs in the kink model with $d^0 = 0.2$ (top panels) and 0.8 (bottom panels) for different choices of $m$ and $n$ in both $\tau_0$ known and unknown cases, for $\sigma = 0.1$ and $0.3$ respectively.} \label{dnotpoint8}
\begin{small}
\begin{minipage}[b]{0.5\linewidth}
\centering
\begin{tabular}{|l||c|c|c|c|}
\hline
$d^0=0.2$ & \multicolumn{3}{c|}{$\sigma=0.1$} \\ \hline
$(m,n)$ & $d^0$ & $d^0 (1),d^0 (2)$ & $\tau_0 (1),\tau_0 (2)$ \\
\hline \hline
(5, 5)   & 0.102 & 0.333, 0.057 & 0.327,  0.048   \\
(5, 10)  & 0.082 & 0.291, 0.060 & 0.260,  0.038  \\
(10, 10)  & 0.081 & 0.259, 0.054 & 0.238,  0.027  \\
(10, 20) & 0.054 & 0.143, 0.045 & 0.127, 0.019   \\
(10, 50)  & 0.031 & 0.044, 0.033 & 0.026, 0.011   \\
(20, 50)  & 0.027 & 0.025, 0.027 & 0.012,  0.008 \\
(50, 100) & 0.015 & 0.015, 0.015 & 0.005, 0.003  \\
\hline
\end{tabular}
\end{minipage}
\hspace{0.5cm}
\begin{minipage}[b]{0.5\linewidth}
\centering
\begin{tabular}{|l||c|c|c|c|}
\hline
$d^0=0.2$ & \multicolumn{3}{c|}{$\sigma=0.3$} \\ \hline
$(m,n)$ & $d^0$ & $d^0 (1),d^0 (2)$ & $\tau_0 (1),\tau_0 (2)$ \\
\hline \hline
(5, 5)  & 0.128 & 0.429, 0.196 & 0.349, 0.138   \\
(5, 10)   & 0.120 & 0.436,  0.181 & 0.329,  0.111  \\
(10, 10)   & 0.095 & 0.396,  0.122 & 0.317,  0.080  \\
(10, 20)  & 0.083 & 0.367, 0.114 & 0.287, 0.058  \\
(10, 50)   & 0.080 & 0.308,  0.112 & 0.229, 0.036   \\
(20, 50)  & 0.060 & 0.210,  0.075 & 0.161,  0.024 \\
(50, 100) & 0.039 & 0.058,  0.045 & 0.024,  0.010   \\
\hline
\end{tabular}
\end{minipage}
\end{small}
\end{table}

\begin{table}
\begin{small}
\begin{minipage}[b]{0.5\linewidth}
\centering
\begin{tabular}{|l||c|c|c|c|}
\hline
$d^0=0.8$ & \multicolumn{3}{c|}{$\sigma=0.1$} \\ \hline
$(m,n)$ & $d^0$ & $d^0 (1),d^0 (2)$ & $\tau_0 (1),\tau_0 (2)$ \\
\hline \hline
(5, 5)   & 0.213 & 0.128, 0.174 & 0.028,  0.028   \\
(5, 10)  & 0.119 & 0.088,  0.118 & 0.023,  0.019  \\
(10, 10)   & 0.126 & 0.095,  0.114 & 0.016,  0.013 \\
(10, 20)   & 0.060  & 0.050,  0.057 & 0.011,  0.008   \\
(10, 50)  & 0.033 &0.033, 0.035 & 0.008,  0.005 \\
(20, 50)   & 0.026 & 0.025, 0.028 & 0.005,  0.004 \\
(50, 100)   & 0.016 & 0.015, 0.015 & 0.002,  0.002 \\
\hline
\end{tabular}
\end{minipage}
\hspace{0.5cm}
\begin{minipage}[b]{0.5\linewidth}
\centering
\begin{tabular}{|l||c|c|c|c|}
\hline
$d^0=0.8$ & \multicolumn{3}{c|}{$\sigma=0.3$} \\ \hline
$(m,n)$ & $d^0$ & $d^0 (1),d^0 (2)$ & $\tau_0 (1),\tau_0 (2)$ \\
\hline \hline
(5, 5)  & 0.178 & 0.114,  0.146 & 0.080,  0.078 \\
(5, 10)   & 0.115 & 0.099,  0.109 & 0.060,  0.051   \\
(10, 10)  & 0.120 & 0.092, 0.111 & 0.044,  0.037   \\
(10, 20)  & 0.087 & 0.088,  0.092 & 0.033,  0.025  \\
(10, 50)   & 0.084 & 0.097,  0.091 & 0.024,  0.016 \\
(20, 50)  & 0.060 & 0.068,  0.064 & 0.017, 0.011  \\
(50, 100)   & 0.038 & 0.042, 0.039 & 0.008,  0.005 \\
\hline
\end{tabular}
\end{minipage}
\end{small}
\end{table}
\noindent {\bf A related allocation problem:} In our simulation study, the proposed procedures for both known and unknown $\tau_0$ were evaluated for a combination of $m$ and $n$ values. However, in practice one is given a {\em total budget} of $N\equiv n\times m$ samples that need to be {\em allocated} to $n$ covariate values and $m$ replicates at each covariate value, respectively. Intuitively, increasing the number of replicates $m$ decreases the ``bias'', whereas increasing the number of values $n$ of the covariate, decreases the variance of the estimators. The optimal allocation occurs when the two terms are balanced, usually at a moderate value of $n$ and $m$ (which depends on the value of $\sigma$ and the regression function). Thus, for a fixed $N$, one expects that the RMSEs exhibit a ``U-shape" as a function of $m$; further, for larger $\sigma$ the optimal allocation would occur at a larger value of $m$.
\newline
\newline
We investigate this allocation problem through a simulation, but due to space considerations we present the optimal allocations for models M1 and M5 for both Methods 1 and 2. The setting under consideration is $d^0=0.5$, $N=100$ and 200 and $\sigma = 0.1$ and $0.3$. All possible combinations of $m$ and $n$ that approximately satisfy the total budget were considered. The optimal allocations are shown in Table \ref{alloc-table}.
\newline
\newline
\begin{table}
\caption{Optimal allocation $(m,n)$ pairs for a fixed total budget $N=m\times n$ \label{alloc-table}}
\centering
\begin{tabular}{ | l || l | cc || l | cc | } \hline
M1 & Method 1 & $\sigma=0.1$ & $\sigma=0.3$ & Method 2 & $\sigma=0.1$ & $\sigma=0.3$ \\ \hline
      &  $N=100$ & (6,17)	&   (33,3) & $N=100$ & (4,25) & (33,	3) \\
      & $N=200$ & (7,29) &  (40,5) & $N = 200$ & (6,33) & (15,13) \\ \hline
 M5 & Method 1 & $\sigma=0.1$ & $\sigma=0.3$ & Method 2 &$ \sigma=0.1$ & $\sigma=0.3$ \\ \hline
 & $N=100$ & (8,12)	  & (33,3) & $N = 100$ & (5,20) & (33,3) \\
       & $N=200$ & (7,29) &  (67,3) & $N = 200$ & (6,33) & (15,13) \\ \hline\hline
\end{tabular}
\end{table}
It can be seen that both methods for small $\sigma$ favor lots of covariate values and few replicates, while the situation is reversed for high $\sigma$. Further, qualitatively similar results, in accordance with our observation above, are obtained for the other three models examined ($M_2$, $M_3$ and $M_4$). Nevertheless, a few anomalies are present; specifically, as we are sampling from the discrete uniform design on $[0,1]$, and $d^0=0.5$, sometimes the optimal allocation occurs at the rather extreme value $n = 3$. This is due to the fact that in that case, the covariate values are placed at 0.25, 0.5 and 0.75, and when $m$ is large, the fitted break point $\hat d_n$ is usually (for many replicates) 0.5, the true parameter value. Whenever this is the case, the estimation error is exactly zero, making the observed RMSEs small. With the same budget, a larger $n$ (say $n = 5$) can also lead to $0.5$ as a covariate value, but the value of $m$ decreases in the process (thereby increasing the bias) and there are more options for the fitted break point to differ from $0.5$, leading to larger RMSEs.
\newline
\newline
{\bf Some practical recommendations:} Based on our extensive simulation study (including results not shown here due to space considerations), the following practical recommendations are in order. Overall, it is better for one to invest in an increased number of covariate values ($n$), rather than replicates ($m$). In the case of a known $\tau$, the simple stump model performs well overall, while the more complicated adaptive stump model offers only marginal improvements. In the case where the threshold $d^0$ is closer to the boundaries, investment in $n$ proves fairly important. For unknown level $\tau_0$, none of the proposed methods dominates the other, the result depending on both the noise level and the model under consideration. However, for estimation of the level $\tau_0$, Method 2 exhibits a clear advantage over its competitor. Some hybrid possibilities are discussed in the concluding remarks section.

\subsection{Comparison with other procedures}
\begin{table}
\caption{RMSEs for the five procedures for different choices of $m$ and $n$ when $\sigma = 0.3$ and the actual model is $M_1$ (left table) and $M_2$ (right table).}\label{CompKinkI}
\begin{small}
\begin{minipage}[b]{0.5\linewidth}\centering
\begin{tabular}{|l||c|c|c|c|c|}
\hline
$(m,n)$ & $P_1$ &  $P_2$ & $P_3$ & $P_4$ & $P_5$\\
\hline \hline
$(5, 5)$& 0.163 &   0.207  &  0.339 &  0.255 &   0.299\\
$(5, 10)$ & 0.134  &  0.176  &  0.304  &  0.307 &   0.344\\
$(10, 10)$ & 0.119  &  0.120  &  0.227   & 0.228 &   0.328\\
$(10, 20)$ & 0.092  &  0.079   & 0.191  &  0.265  &  0.295\\
$(10, 50)$ & 0.085  &  0.042  &  0.179  &  0.310 &   0.247\\
$(20, 50)$ & 0.060 &   0.030 &   0.128 &   0.212 &   0.176\\
$(50, 100)$ & 0.038  &  0.013 &   0.080 &   0.142  &  0.114\\
\hline
\end{tabular}
\end{minipage}
\hspace{0.05cm}
\begin{minipage}[b]{0.5\linewidth}
\centering
\begin{tabular}{|c||c|c|c||c|}
\hline
$P_1$ &  $P_2$ & $P_3$ & $P_4$ & $P_5$\\
\hline
0.204  &  0.241 &   0.420 &   0.291 &   0.298\\
0.201  &  0.227  &  0.390 &   0.346  &  0.360\\
0.168  &  0.194 &   0.334 &   0.302  &  0.360\\
0.177  &  0.163 &   0.303  &  0.329 &   0.354\\
0.193  &  0.150 &   0.294 &   0.369  &  0.332\\
0.162  &  0.147 &   0.245  &  0.305  &  0.274\\
0.132  &  0.145  &  0.197  &  0.254  &  0.211\\
\hline
\end{tabular}
\end{minipage}
\end{small}
\end{table}

\begin{table}
\caption{RMSEs for the five procedures for different choices of $m$ and $n$ when $\sigma = 0.3$ and the actual model is $M_3$ (left table) and $M_5$ (right table).}\label{CompKinkII}
\begin{small}
\begin{minipage}[b]{0.5\linewidth}\centering
\begin{tabular}{|l||c|c|c|c|c|}
\hline
$(m,n)$ & $P_1$ &  $P_2$ & $P_3$ & $P_4$ & $P_5$\\
\hline \hline
$(5, 5)$& 0.173  &  0.197 &   0.342  &  0.245  &  0.314\\
$(5, 10)$ & 0.140 &   0.159 &   0.306 &   0.297 &   0.351\\
$(10, 10)$ & 0.126 &   0.116 &   0.247  &  0.228  &  0.319\\
$(10, 20)$ & 0.117  &  0.084  &  0.216 &   0.256  &  0.282\\
$(10, 50)$ & 0.129  &  0.068  &  0.203  &  0.302  &  0.248\\
$(20, 50)$ & 0.110  &  0.064   & 0.170 &   0.213  &  0.194\\
$(50, 100)$ & 0.098  &  0.060  &  0.138  &  0.164  &  0.151\\
\hline
\end{tabular}
\end{minipage}
\hspace{0.05cm}
\begin{minipage}[b]{0.5\linewidth}
\centering
\begin{tabular}{|c||c|c|c||c|}
\hline
$P_1$ &  $P_2$ & $P_3$ & $P_4$ & $P_5$\\
\hline
0.181 &   0.232  &  0.376  &  0.277 &   0.286\\
0.153  &  0.239  &  0.370 &   0.370 &   0.287\\
0.117  &  0.203  &  0.240  &  0.336  &  0.282\\
0.093  &  0.191  &  0.198  &  0.411  &  0.266\\
0.084  &  0.168  &  0.178 &   0.465  &  0.241\\
0.060  &  0.148  &  0.127   & 0.440   & 0.175\\
0.038   & 0.139 &   0.080  &  0.402   & 0.113\\
\hline
\end{tabular}
\end{minipage}
\end{small}
\end{table}
Next, we compare the proposed 1--parameter stump method to some competing procedures developed in the pharmacological dose--response setting to identify the MED. Most of the methods developed in dose--response setting context are based on hypothesis testing procedures. For example, Williams (1971) developed a method to identify the lowest dose at which there is ``activity'' in toxicity studies using a closed testing procedure based on isotonic regression for a monotone dose--response relationship. Hsu and Berger (1999) developed a step--wise confidence set approach to estimate and make inference on the MED. A nonparametric method based on the Mann-Whitney statistic incorporating the step--down procedure is investigated in Chen (1999), while Tamhane and Logan (2002) use multiple testing procedures for the task at hand. We compare our method with that of Williams (1971), of Hsu and Berger (1999) and of Chen (1999), referred henceforth as $P_3, P_4$ and $P_5$, respectively.
\newline
\newline
We fit the stump $\xi_d$ with levels 1/2 and 0 on either side of the threshold $d$ to the observed $p$--values (with $\tau_0 \equiv 0$ assumed known) and compare the performance of our approach $P_1$ with that of $P_3$, $P_4$ and $P_5$. A natural parametric procedure to estimate $d^0$ might be to fit a kink--type (hockey stick) model like $M_1$ to the observed responses and estimate $d$ (the threshold parameter) and the slope of the linear segment by the least squares method. We also implement this method and call it $P_2$. Obviously when the true underlying regression function $\mu$ is not a kink--model this method might not be consistent, but given a finite sample it is often a good first approximation. Whereas, when $\mu$ is a kink--function, e.g., when we assume the true model to be $M_1$, this approach should clearly outperform the other procedures. Indeed, Table 5 shows that $P_2$ is very competitive for model $M_1$; still our approach $P_1$ performs better for small sample sizes, e.g., (5, 5) and (5, 10). For the model $M_2$, a slight departure from the model $M_1$, $P_1$ mostly dominates $P_2$, and all the other procedures. Note that as $P_3$, $P_4$ and $P_5$ are procedures that are based on testing hypotheses, we need to specify a level ($\alpha$), and in the simulations reported in the paper we have set $\alpha = 0.05$. The choice of the $\alpha = 0.05$ is purely based on classical hypothesis testing considerations; a proper choice of the tuning parameter is not available. Changing $\alpha$ will change the RMSEs of $P_3$--$P_5$, and it is not quite clear how that will affect the estimation procedure and the RMSEs. Also, to implement $P_3$--$P_5$, we computed the cut-off values necessary to carry out the hypothesis tests using simulation, as such tables are not available for the different choices of $m$ and $n$ considered in this paper.
\newline
\newline
Table 6 shows the performance of the methods when the true models are $M_3$, the infinitely differentiable regression function, and $M_5$. Clearly $P_1$ dominates all the other methods when the data is generated according to $M_5$; notice that $P_3$ and $P_4$ work with the underlying assumption that $\mu$ is nondecreasing, a condition which is violated in $M_5$, and this explains the poor performance of these methods. Also, $P_2$ is biased in this scenario, and thus the RMSEs do not converge to 0 for large $m$ and $n$. When $M_3$ is the true model, $P_2$ performs surprisingly well; this can be explained by looking at Figure \ref{six-reg-funcs}, the threshold value for $M_3$ is approximated very well by that of the kink--model. This is an artifact of the particular choice of the infinitely differentiable function, and the RMSEs can in general be very different if the $\mu$ is not well approximated by a kink function. Overall, $P_1$ is very competitive, and the simplicity of our approach coupled with its adaptivity to different types of mis--specifications, makes it a very attractive choice. We also note that the fitting of $\xi_d$ does not require any tuning parameter, an obvious advantage over the testing based procedures. Indeed, one of the novelties of our approach lies in the fact that we treat the estimation of $d^0$ purely as an estimation problem and not a result of a series of hypotheses tests, thereby avoiding the need to specify $\alpha$.
\subsection{Data Applications}\label{data-application}
\begin{figure}[!h]
\centering
\resizebox{5.5in}{2in}{\includegraphics{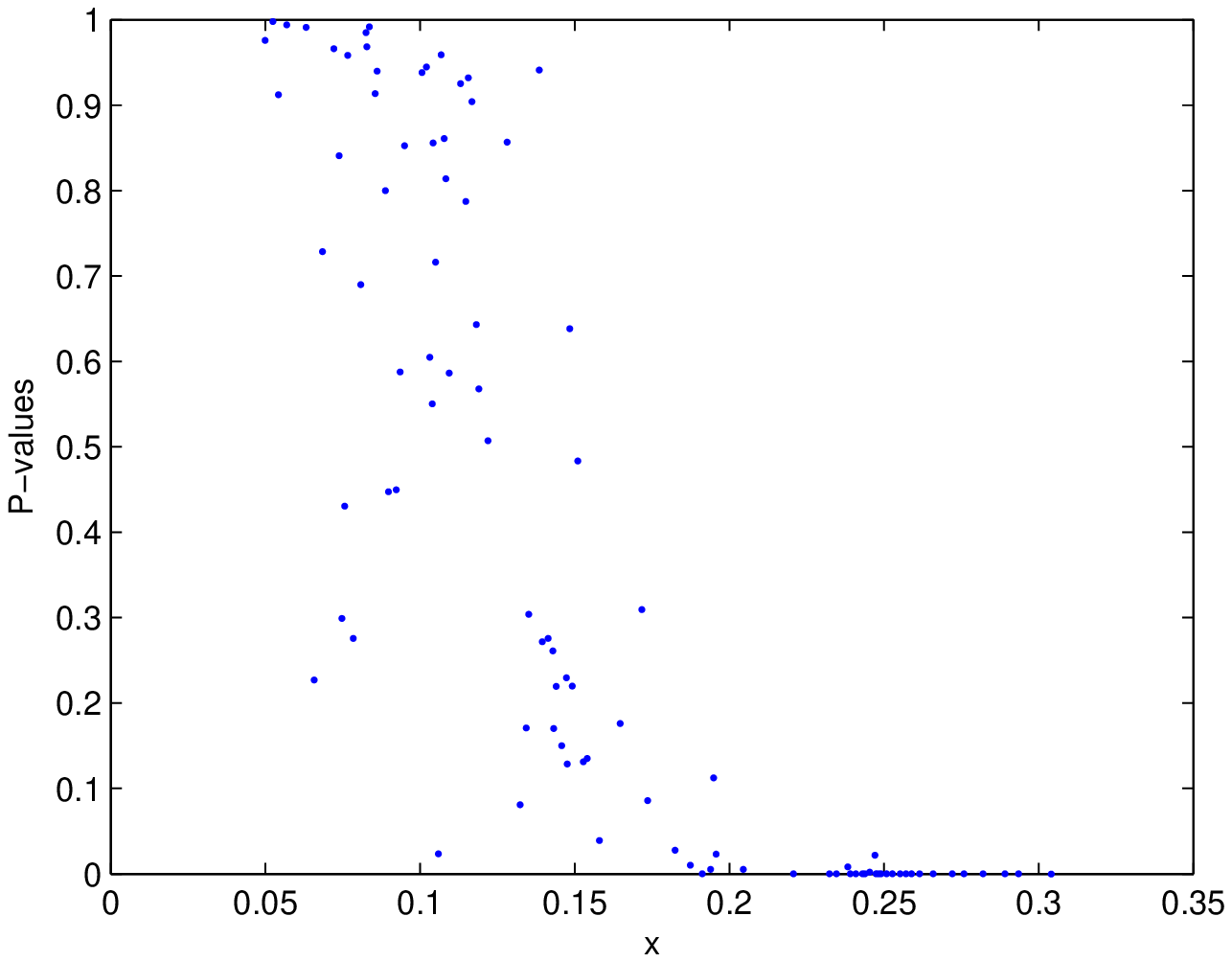} \ \
\includegraphics{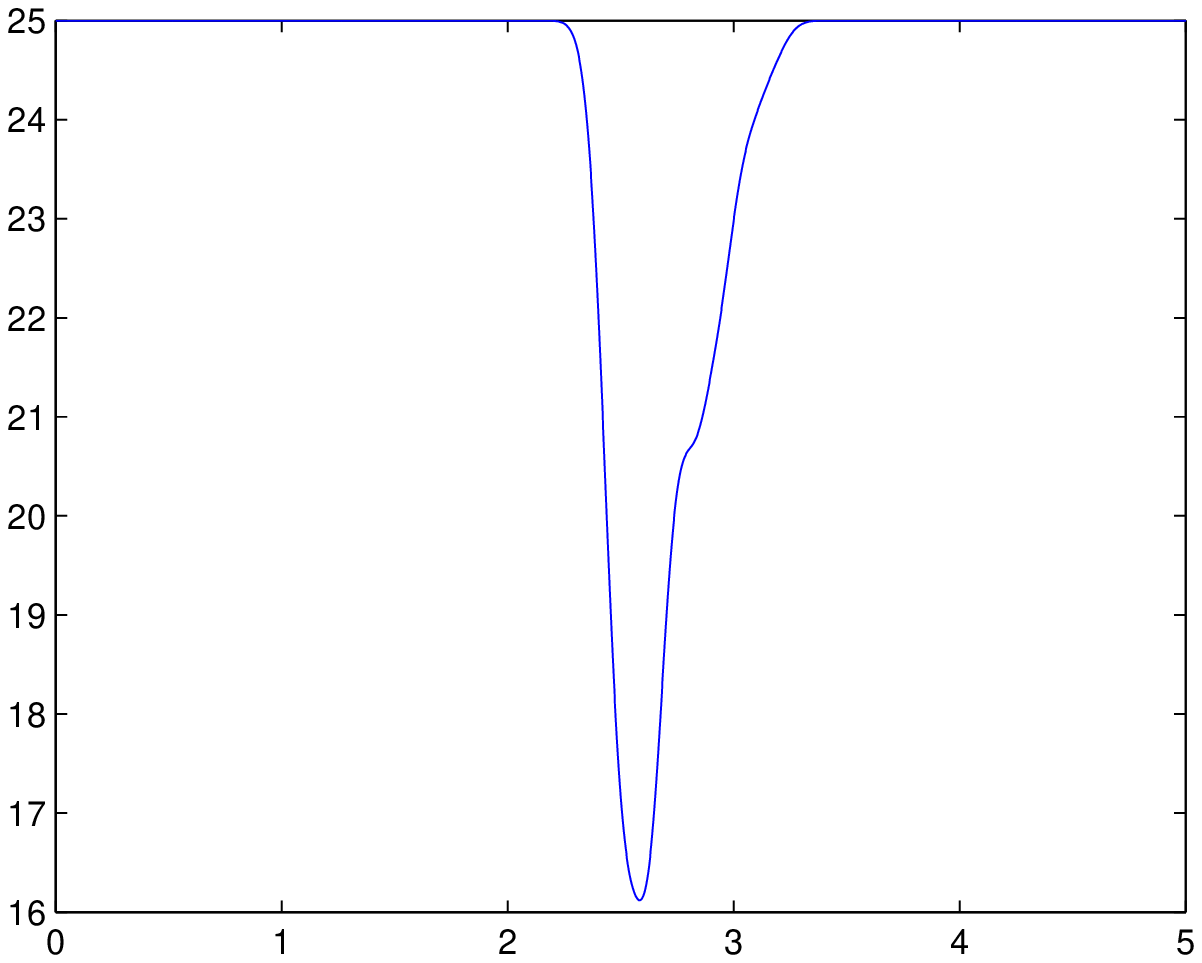}}
\caption{Plots of the estimated $p$--values using Method 1 (left panel) and the plot of the criterion function whose minimizer provides the estimate of $\tau_0$ (right panel).}\label{pvalue-crite-plot}
\end{figure}

\begin{figure}[!h]
\centering
\resizebox{2.5in}{1.5in}{\includegraphics{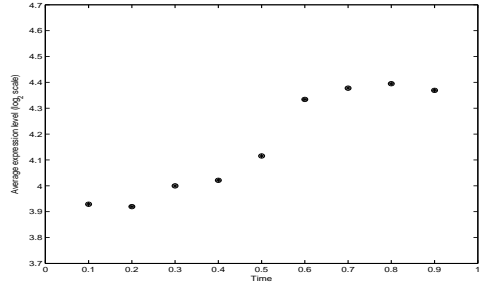}}
\caption{Average expression of the Transglutaminase 2 gene over time. \label{ave-expression}}
\end{figure}

In this section, we apply the proposed procedure to the two motivating applications. Note that the first one corresponds to a rich allocation scheme in terms of $(m,n)$, while the second application to a sparse one, thus showing the range of applicability of the procedure. Further, for the first application we discuss a subsampling mechanism that allows us to calculate confidence intervals for $\hat{d}_0$, given the large number of doses available. This is not repeated for the gene expression data due to the paucity of time points, but for richer time course experiments it would become applicable.
\newline
\newline
{\bf Queueing System:}
We consider a complex system comprising multiple classes of customers waiting at infinity capacity queues and a set of processing resources modulated by an external stochastic process. The system employs a resource allocation (scheduling) policy that decides at every time slot which customer class to serve, given the state of the modulating rate process and the backlog of the various queues. In Bambos and Michailidis (2004), a low complexity policy was introduced and its maximum throughput properties established. This canonical system captures the essential features of data/voice transmissions in a wireless network, in multi-product manufacturing systems, and in call centers (for more details see Bambos and Michailidis (2004)). As discussed in the introductory section, an important quantity of interest to the system's operator is the average delay of jobs (over all classes), which constitutes a key performance metric of the quality of service offered by the system. The average delay of the jobs in a two-class system as a function of its loading under the optimal policy, for a small set of loadings is shown in the left panel of Figure \ref{full-data}. These responses were obtained through simulation, since for such complex systems analytic calculations are intractable.
\begin{figure}[!h]
\centering
\resizebox{4in}{2.5in}{\includegraphics{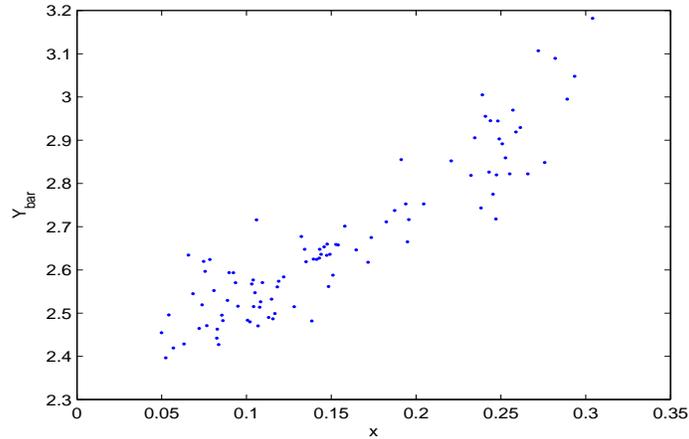}}
\caption{\label{Ybar-fig} Plot of the average responses $\bar{Y}_i$.}
\end{figure}
\newline
\newline
We next employ the developed methodology for estimating both the loading $d^0$ and the unknown level $\tau_0$. Ten replicates of the response (average delay) were obtained based on 5,000 events per class by simulating the system under consideration and after accounting for a burn-in period of 2,000 per class in order to ensure that it reached its stationary regime. The means per loading, $\bar{Y}_i$s, are shown in Figure \ref{Ybar-fig}.
\newline
\newline
We applied both procedures discussed in Section \ref{sec:unknown-tau} with an unknown value for $\tau_0$, assuming heteroscedastic errors. The first method (Method 1) gives an estimate of $\tau_0$, denoted by $\hat{\tau}_0(1) = 2.61$ with corresponding $\hat{d}^0(1) = 0.151$, while the second method (Method 2) gives $\hat{\tau}_0(2)=2.52$ with corresponding $\hat{d}^0(2)=0.117$. It can be seen that there is fairly strong agreement between the two estimates. From the system's operator point of view the average delay of jobs exhibits a markedly increasing trend beyond a loading of 15\%. For the first method, a plot of the P-values (left panel) and the criterion function (right panel) that is minimized in Theorem 3.2 are given in Figure \ref{pvalue-crite-plot}.
\newline
\newline
A second analysis, assuming homoscedastic errors, produces fairly comparable results: $\hat{d}^0(1)=0.151$ and $\hat{\tau}_0(1)=2.59$ and $\hat{d}^0(2)=0.128$ and $\hat{\tau}_0(2)=2.52$, respectively for the two methods.
Another question of interest to the system's operation is what level of uncertainty, as reflected through confidence intervals, can be assigned to these estimates. Obviously, our results establish consistency of the threshold $\hat{d}^0$ and level $\hat{\tau}_0$ estimates, but no characterization of their asymptotic distribution has been provided, which would have resolved this issue. Nevertheless, we outline a subsampling based procedure for partially addressing the construction of confidence intervals, provide some theoretical justification in the Appendix and finally discuss some open issues (see Section \ref{sec:conclusions}).
\newline
\newline
The steps in the employed procedure are:
\begin{enumerate}
\item Sample $m_n$ vectors out of the 100 $\{X_i,\bar{Y}_i\}$ pairs without replacement.

\item For the collected sub-sample, compute estimates of $d^0$ via the two methods, denoted by $\hat{d}^{0,\star}(1)$ and $\hat{d}^{0,\star}(2)$ respectively. Two versions of $\hat{d}^{0,\star}(1)$ are calculated: the first, denoted by $\hat{d}^{0,\star}(1,a)$ is calculated using the estimate of $\tau_0$ based on the full sample (in other words, taking $\hat{\tau}_0(1)$ as the ``truth"), while the second denoted by $\hat{d}^{0,\star}(1,b)$ is computed after re-estimating $\tau_0$ from the obtained subsample.

\item Calculate the following statistics: $t_{1n}^{\star} \equiv m_n^{1/3}\,(\hat{d}^{0,\star}(1,a) - \hat{d}^{0}(1)), t_{2n}^{\star} \equiv m_n^{1/3}\,(\hat{d}^{0,\star}(1,b) - \hat{d}^{0}(1))$ and $t_{3n}^{\star} \equiv m_n^{1/3}\,(\hat{d}^{0,\star}(2) - \hat{d}^{0}(2))\,$.

\item Repeat the above 3 steps a large number of times (say $B$), storing the three statistics in the preceding step for each iteration, and obtain the empirical distributions for each of these statistics based on the $B$ iterates, say $\{F_{j,n}^{\star}\}_{j=1}^3$.

\item Calculate $q_{.025,j,n}^{\star}$ and $q_{.975,j,n}^{\star}$, the 2.5-th and 97.5-th percentiles of $F_{j,n}^{\star}$ respectively, for $j = 1,2,3$.

\item Prescribe $[\hat{d}^{0}(1) - n^{-1/3}q_{.975,j,n}^{\star}, \hat{d}^{0}(1) - n^{-1/3}q_{.025,j,n}^{\star}]$, for $j = 1,2$, $\mbox{ and }$ \\ $[\hat{d}^{0}(2) - n^{-1/3}q_{.975,3,n}^{\star}, \hat{d}^{0}(2) - n^{-1/3}q_{.025,3,n}^{\star}]$ as approximate 95\% confidence intervals for $d^0$.
\end{enumerate}
Using $m_n=50$ and assuming heteroscedastic errors the 95\% confidence intervals for $d^0$ when $j=1,2,3$, are (.133, .179),  (.135,.158) and (.099,.125) respectively. Very similar results are obtained for $m_n=75$, while for smaller values of $m_n$ (e.g. 10 or 25) the resulting confidence intervals become exceedingly wide. Under the assumption of homoscedasticity the obtained 95\% confidence intervals are (0.133, 0.178), (0.151, 0.176) and (0.120, 0.139), respectively. It should be noted that the third confidence interval does not overlap with the second one, something that can be attributed to their fairly narrow respective lengths. From the system's operator perspective, it can be seen that running the system at loadings below 12\% of the total capacity produces very small delays on the average for its customers, while at loadings larger than 18\%, customers should expect to experience increasing delays.
\newline
\newline
{\bf Time course analysis of the Transglutaminase 2 gene:} This example deals with a time course experiment that studies the effects of cells treated with TGF-$\beta$, a key cytokine implicated in a number of disease processes including cancer, on the epithelial-mesenchymal transition, a phenotypic conversion that enables cancer cells to attain their migratory and invasive capacities (see Keshamouni et al. (2009)). The data correspond to the expression levels of the Transglutaminase 2 gene measured in triplicate at 0, 0.5, 1, 2, 4, 8, 16, 24 and 72 hours. This gene has been implicated in this mechanism through an elevated level at 72 hours, compared to the baseline at 0 hours (see Keshamouni et al. (2006)). The questions of interest are (a) at what time point its expression level rises from its baseline, since this provides a timeline for the activation of this transition mechanism in cell-lines; (b) at what time point its expression level reaches saturation, which indicates that transition to malignancy is essentially complete.
\newline
\newline
The average expression level of the 3 replicates is shown in Figure \ref{ave-expression} and it can be seen that it rises from its initial baseline level and subsequently flattens out. It is therefore reasonable to postulate a model that is constant ($\tau_{\min}$) till the first transition point $d_{\min}$, then increases monotonically until it reaches the second transition point $d_{\max}$ beyond which it remains constant at $\tau_{\max}$.
Without loss of generality, for the analysis, the nine design points were taken as equispaced in the $[0,1]$ interval, since our interest focuses on identifying the stages where the changes occur. The procedure described in Section \ref{multiple-pts} was employed to estimate the parameters of interest with the $p$--values being computed under the assumption of heteroscedasticity of the error (as evident from the right panel of Figure \ref{full-data}). For the baseline, Method 1 gives an estimate $\hat{\tau}_{\min}(1)=3.974$ with corresponding $\hat{d}_{\min}(1)=0.4$, while Method 2 gives $\hat{\tau}_{\min}(2)=3.924$ with corresponding $\hat{d}_{\min}(2)=0.2$. It can be seen that the two methods basically agree on the value of $\tau_{\min}$, but the second one identifies the change-point somewhat earlier than the first method. For the maximum level, Method 1 gives an estimate $\hat{\tau}_{\max}(1)=4.367$ with corresponding $\hat{d}_{\max}(1)=0.6$, while Method 2 gives $\hat{\tau}_{\max}(2)=4.369$ with corresponding $\hat{d}^{\max}(2)=0.6$, thus exhibiting perfect agreement. To estimate the minimum (baseline) value using Method 1 we minimized (\ref{eq:Tau_mn}) over the restricted interval $[\min(\bar Y_{i.}), 4]$ and for the maximum over the restricted interval $(4, \max(\bar Y_{i.})]$ respectively, as advocated in Section \ref{multiple-pts}. The choice of the intervals do not matter so long they are disjoint and are contain the true minimum and maximum. We believe that the 0.4 time point is a more accurate estimate of $d_{\min}$, since the proteins encoded by the Transglutaminase 2 gene exhibit increased levels at 8 hours (the 0.5 design point) and nothing significant prior to that time, as obtained from a separate experimental platform described in Keshamouni et al. (2009).

\section{Concluding Discussion} \label{sec:conclusions}
In this paper, we address the problem of identifying the threshold parameter at which a regression function diverges from its (possibly unknown) baseline value employing a $p$--value framework, under a controlled sampling setting. The $p$--values exhibit a natural dichotomy in behavior on different sides of the threshold, which is crucially used in our methodology. Our approach is computationally simple, nonparametric in nature and adaptive in the sense that it does not need any specification of the local behavior of the regression function around the threshold value. The procedure can also be used to prescribe estimates for the baseline value of the regression function. We establish asymptotic properties of the proposed estimators, extend them to the case of multiple change-points, study their finite sample behavior through an extensive simulation study, compare them for the MED problem to competing approaches and apply them to two real applications. The numerical results indicate that the procedure performs well in both high and relatively low signal--to--noise settings.
\newline
\newline
We conclude with a discussion of a number of issues, some of which will be the focus of future work. In our setting, we dealt with the case of a balanced design with a fixed number of replicates $m$ for every dose level $X_i$. The case of varying number of replicates $m_i$ can be handled analogously, although some care needs to be exercised in the technical arguments. Under the assumption that the minimum of the $m_i$'s goes to infinity, all consistency results in this paper continue to hold. A natural extension of the problem would be to investigate the performance of our method when we have a random design of points with no replicates, i.e., $m_i = 1$. Such problems arise in diverse contexts, e.g., in the astronomy application discussed in the Introduction, where the goal is to estimate the ``tidal'' radius. Note that in this setup, (approximate) $p$--values can be computed by binning the covariate space and averaging the responses in each bin.
\newline
\newline
The problem of constructing confidence intervals for the threshold $d^0$ is of considerable interest in many applications. \emph{Ideally}, one would like to consider the asymptotic distribution of $\hat{d}_{m,n}$, as both $m, n$ increase to infinity and use the quantiles of the resulting distribution to calibrate confidence intervals. However, as discussed in the previous section, this is a hard problem and its solution, presently unknown, is outside the scope of this paper. This constitutes a subject of future research.
\newline
\newline
In the case of an unknown threshold $\tau_0$, our numerical results show a superior performance of Method 2 over its competitor, especially in small $(m,n)$ scenarios. The poorer performance of Method 1 in these situations was tracked down to the inaccurate estimation of $\tau_0$ in the simulation section. Note that consistency of the estimator is formally established for Method 1. It would therefore be interesting to explore a hybrid procedure and its properties, where $\tau_0$ is estimated using Method 2 and the resulting estimate is used as a plug-in in Method 1.
\newline
\newline
Although we develop our method in a simple univariate regression setup, our approach, can be generalized to identify the ``baseline'' region in multi-dimensional covariate spaces. Further, for regression models in higher dimensions, the problem of estimating regions in the covariate space where the regression function stays below a pre-specified threshold can also be handled by our approach. Such problems, known as \emph{level-sets estimation}, have been extensively studied in the statistics and engineering literature (see for example, Singh et al. (2009), Willet and Nowak (2007)). Indeed, Section \ref{comp-hyp} of the current paper may be viewed as a level-set estimation problem, albeit in a simpler setting, but under minimal assumptions on the behavior of the regression function at the boundary.
\newline
\newline
In conclusion we recall that the problem treated in this paper could also have been treated by direct estimation of the underlying regression function $\mu$. However, as briefly mentioned in the Introduction, straightforward intuitive estimates of the form $\hat d \equiv \inf\{x: \hat\mu(x)>0\}$ underestimates $d^0$; potential fixes lead to solutions for which very little is known in terms of asymptotic properties. Nevertheless, it would be fruitful to explore that line of research.
\newline
\newline
{\bf Acknowledgements} \\ \\ We would like to thank Harsh Jain for bringing to our attention a threshold estimation problem that eventually led to the formulation and development of this framework. The work of the authors were partially supported by NSF and NIH grants: DMS-09-06597 (BS), DMS-07-05288 (MB) and 1RC1CA145444-0110 (GM).
\section{Appendix}
We start with establishing an auxiliary result used in subsequent developments.
\begin{theorem}
\label{uniform-cts-mapping}
Let $\{\mathbb{M}_n^{\tau}: \tau \in \mathcal{T}\}_{n=1}^{\infty}$ be a family of (real-valued) stochastic processes indexed by $h \in \mathcal{H}$ and let $\{M^{\tau}: \tau \in \mathcal{T}\}$ be a family of deterministic functions defined on $\mathcal{H}$, such that each $M^{\tau}$ is minimized at a unique point $h(\tau) \in \mathcal{H}$. Here $\mathcal{H}$ is a metric space and denote the metric on $\mathcal{H}$ by $d$. Let $\hat{h}_n^{\tau}$ be a minimizer of $\mathbb{M}_n^{\tau}$. Assume further that:
\newline
\newline
(a) $\sup_{\tau \in \mathcal{T}}\,\sup_{h \in \mathcal{H}}\,|\mathbb{M}_n^{\tau}(h) - M^{\tau}(h)| \stackrel{p}{\rightarrow}  0 $ and \newline
(b) For every $\eta > 0$, $c(\eta) \equiv  \inf_{\tau}\,\inf_{h \notin B_{\eta}(h(\tau))}\,\{M^{\tau}(h) - M^{\tau}(h^{\tau}) \} > 0$, where $B_{\eta}(h)$ denotes the open ball of radius $\eta$ around $h$.
\newline
\newline
Then, (i) $\sup_{\tau}\,d(\hat{h}_n^{\tau}, h^{\tau})$ converges in probability to 0. Furthermore, if $\mathcal{T}$ is a metric space and $h^\tau$ is continuous in $\tau$, then (ii) $\hat{h}_n^{\tau_n} \stackrel{p}{\rightarrow} \,h^{\tau_0}$, provided $\tau_n$ converges to $\tau_0$. In particular, if the $\mathbb{M}_n^{\tau}$'s themselves are deterministic functions, the conclusions of the theorem hold with the convergences in probability in (i) and (ii) replaced by usual non-stochastic convergence.
\end{theorem}
{\bf Proof:} We provide the proof in the case that $\mathcal{H}$ is a sub-interval of the real line, the case that is relevant for our applications. However, there is no essential difference in generalizing to the metric space case. Euclidean distances simply need to be replaced by the metric space distance and open intervals by open balls.
\newline
Given $\eta > 0$, we need to deal with $P^{\star}\,\{\sup_{\tau \in \mathcal{T}}\,|\hat h_n^{\tau} - h(\tau)| > \eta\}$. We deal with outer probabilities to avoid measurability difficulties. The event $A_{n,\eta} \equiv \{\sup_{\tau \in \mathcal{T}}\,|\hat h_n^{\tau} - h(\tau)| > \eta\}$ implies that for some $\tau$, $\hat h_n^{\tau} \notin (h(\tau) - \eta, h(\tau) + \eta)$ and therefore \[ M^{\tau}(\hat h_n^{\tau}) - M^{\tau}(h(\tau)) \geq \inf_{h \notin (h(\tau) - \eta, h(\tau) + \eta)}\,\{M^{\tau}(h) - M^{\tau}(h(\tau))\} \,.\]
This is equivalent to \[ M^{\tau}(\hat h_n^{\tau}) - M^{\tau}(h(\tau)) - \mathbb{M}_n^{\tau}(\hat h_n^{\tau}) + \mathbb{M}_n^{\tau}(h(\tau)) \geq \] \[ \qquad \qquad \qquad \qquad \qquad \qquad \inf_{h \notin (h(\tau) - \eta, h(\tau) + \eta)}\,\{M^{\tau}(h) - M^{\tau}(h(\tau))\} + \mathbb{M}_n^{\tau}(h(\tau)) - \mathbb{M}_n^{\tau}(\hat h_n^{\tau}) \,.\]
Now, $\mathbb{M}_n^{\tau}(h(\tau))- \mathbb{M}_n^{\tau}(\hat h_n^{\tau}) \geq 0$ and the left side of the above display is dominated by \[2\,\|\mathbb{M}_n^{\tau} - M^{\tau}\|_{\mathcal{H}} \equiv 2\,\sup_{h \in \mathcal{H}}\,|\mathbb{M}_n^{\tau}(h) - M^{\tau}(h)| \,,\]
implying that:
\[ 2 \|\mathbb{M}_n^{\tau} - M^{\tau}\|_{\mathcal{H}} \geq \inf_{h \notin (h(\tau) - \eta, h(\tau) + \eta)}\,\{M^{\tau}(h) - M^{\tau}(h(\tau))\} \,,\]
which, in turn, implies that: \[  2\,\sup_{\tau \in \mathcal{T}}\|\mathbb{M}_n^{\tau} - M^{\tau}\|_{\mathcal{H}} \geq \inf_{\tau \in \mathcal{T}}\,\inf_{h \notin (h(\tau) - \eta, h(\tau) + \eta)}\,\{M^{\tau}(h) - M^{\tau}(h(\tau))\} \equiv c(\eta)\,,\]
by definition. Hence \[ A_{n,\eta} \subset \{\sup_{\tau \in \mathcal{T}}\|\mathbb{M}_n^{\tau} - M^{\tau}\|_{\mathcal{H}} \geq c(\eta)/2 \} \,.\] By assumptions (a) and (b), $P^{\star}\,\{\sup_{\tau \in \mathcal{T}}\|\mathbb{M}_n^{\tau} - M^{\tau}\|_{\mathcal{H}} \geq c(\eta)/2\}$ goes to 0 and therefore so does $P^{\star}(A_{n,\eta})$. $\Box$ \newline

{\bf Remarks:}  We will call the sequence of steps involved in deducing the inclusion: \[ \left\{ \sup_{\tau \in \mathcal{T}}\,|\hat h_n^{\tau} - h(\tau)| > \eta \right\} \subset \left\{ \sup_{\tau \in \mathcal{T}} \|\mathbb{M}_n^{\tau} - M^{\tau}\|_{\mathcal{H}} \geq c(\eta)/2 \right\} \,,\]
as \emph{generic steps}. Very similar steps will be required time and again at places in the proofs of the theorems to follow. We will not elaborate those arguments, but refer back to the \emph{generic steps} in such cases.
\newline
\newline
{\bf Proof of Theorem \ref{consistency--theorem}:} We prove Part (b) of the theorem since Part (a) follows by an (easier) adaptation of the arguments needed for Part (b). Recall that in Part (b), we find the best fitting stump to the observed $p$--values $Z_{im}, i = 1,2,\ldots,n$. Letting $\xi_\theta(x) \equiv  \alpha {1}(x \le d) + \beta {1}(x > d)$ for $\theta = (\alpha,\beta,d)$, we minimize
\begin{eqnarray}\label{eq:CPSSE2}
\mathbb{M}_{m,n}(\theta) = \sum_{i=1}^n \{ Z_{im} - \xi_\theta(X_i)\}^2 =
\sum_{i: X_i \le d} (Z_{im} - \alpha)^2 +  \sum_{i: X_i > d} (Z_{im} - \beta)^2
\end{eqnarray}
over $\theta= (\alpha,\beta,d) \in [0,1]^3$. Letting $\hat \theta_{m,n} = (\hat \alpha_{m,n}, \hat \beta_{m,n}, \hat d_{m,n}) \equiv  \arg \min_{\theta \in [0,1]^3} \mathbb{M}_{m,n}(\theta)$, we see that $\hat \theta_{m,n}$ is a natural estimator of $\theta_0 \equiv  (0.5, 0, d^0)$. Let $\| \cdot \|$ denote the $l_{\infty}$-metric in $\mathbb{R}^3$, i.e, $\| (a,b,d) \| \equiv  \max \{|a|,|b|,|d|\}$. Note that as $m$ changes, the distribution of $Z_{im}$ changes, and so we effectively have a triangular array of i.i.d. random variables $\{(X_i,Z_{im}) \}_{i=1}^n \sim P_m$. It suffices to show that $\hat \theta_{m,n} \stackrel{p}{\rightarrow} \theta_0$ as $m,n {\rightarrow} \infty$, i.e., given $\epsilon, \xi > 0$, there exists $K \in \mathbb{N}$, such that for all $m,n \ge K$, $P_m \{ \|\hat \theta_{m,n} - \theta_0 \| > \epsilon\} < \xi$.
\newline
\newline
Using empirical process notation, note that $\mathbb{M}_{m,n}(\theta) \equiv \mathbb{P}_{n,m} \{Z_{1m} - \xi_\theta(X_1)\}^2$ and define $M_m(\theta) \equiv  P_m \{Z_{1m} - \xi_\theta(X_1)\}^2$ where $M_m(\theta)$ can be simplified as
\begin{eqnarray}\label{eq:defProc}
M_m(\theta) = \int_0^d \{\nu_m(x) - \alpha\}^2 p_X(x) dx + \int_d^1 \{\nu_m(x) - \beta \}^2 p_X(x) dx + c_m,
\end{eqnarray}
with $c_m = \int_0^1 \sigma_m^2(x) p_X(x) dx$, where $\sigma_m^2(x) =$Var$(Z_{im}|X_i = x)$. Note that $\sigma_m^2(x) \rightarrow (1/12){1} (x \le d^0) \equiv \sigma^2(x)$ as $m \rightarrow \infty$. Let $M(\theta)$ be the same expression for $M_m(\theta)$ in (\ref{eq:defProc}) with $\nu_m(x)$ replaced by $\nu(x) = (1/2) {1} (x \le d^0)$ and $c_m$ replaced by $c = \int_0^1 \sigma^2(x) p_X(x) dx$, e.g., $M(\theta) = \int_0^d \{\nu(x) - \alpha\}^2 p_X(x) dx + \int_d^1 \{\nu(x) - \beta \}^2 p_X(x) dx + c$. Observe that $M(\theta) \ge c$ for all $\theta$, and $M(\theta_0) = c$. Also it is easy to observe that $\theta_0$ is the unique minimizer of $M(\theta)$. Note that $M_m(\theta)$ can be expanded as
\begin{eqnarray}\label{eq:M_m}
M_m(\theta) & = & \int_0^1 \nu_m^2(x) p_X(x) dx - 2 \alpha \int_0^d \nu_m(x) p_X(x) dx \nonumber \\
& - & 2 \beta \int_d^1 \nu_m(x) p_X(x) dx + \alpha^2 \int_0^d p_X(x) dx + \beta^2 \int_d^1 p_X(x) dx + c_m. \nonumber
\end{eqnarray}
Using a similar expansion for $M(\theta)$, the difference $|M_m(\theta) - M(\theta)|$, can be bounded as
\begin{eqnarray}
& & 2 \int_0^1 |\nu_m^2(x) - \nu^2(x)| p_X(x) dx + 2 \alpha \int_0^d |\nu_m(x) - \nu(x)| p_X(x) dx \nonumber \\
& - & 2 \beta \int_d^1 |\nu_m(x) - \nu(x)| p_X(x) dx + |c_m - c| \nonumber \\
& \le & 8 \int_0^1 |\nu_m(x) - \nu(x)| p_X(x) dx + |c_m - c| \rightarrow 0 \nonumber
\end{eqnarray}
uniformly in $\theta \in [0,1]^3$, e.g., $\|M_m - M\|_{\infty} \equiv  \sup_{\theta \in [0,1]^3} |M_m(\theta) - M(\theta)| \rightarrow 0$. By Theorem \ref{uniform-cts-mapping}, $\theta_{m} = (\alpha_m,\beta_m,d_m) = \arg \min_{\theta \in [0,1]^3} M_m(\theta) \rightarrow \arg \min_{\theta \in [0,1]^3} M(\theta) = \theta_0$ as $m \rightarrow \infty$.

Notice now that it is enough to show that for any $\epsilon> 0$, for some $M_0$ (possibly depending on $\epsilon$),
\begin{eqnarray}\label{eq:Cons_d_mn}
\sup_{m \ge M_0} P_m\{\sup_{n \ge k} \| \hat \theta_{m,n} - \theta_m \| > \epsilon \} \rightarrow 0, \mbox{ as } k \rightarrow \infty.
\end{eqnarray}
To see this, take any $\xi >0$. Then, by (\ref{eq:Cons_d_mn}) and the fact that $\theta_m \rightarrow \theta_0$, there exists $K \in \mathbb{N}, K > M_0$ such that for all $k \ge K$, \[\sup_{m \ge M_0} P\{\sup_{n \ge k} \|\hat \theta_{m,n} - \theta_m\| > \epsilon/2 \} \le \xi, \mbox{ and } \|\theta_k - \theta_0\| < \epsilon/2, \] which implies for $n,m \ge K$, \[P\{\|\hat \theta_{m,n} - \theta_0\| > \epsilon\} \le P\{ \|\hat \theta_{m,n} - \theta_m \| > \epsilon/2 \} \le P\{\sup_{n \ge K} \|\hat \theta_{m,n} - \theta_m \| > \epsilon/2\} \le \xi,\] thereby completing the argument.
\newline
\newline
To show that (\ref{eq:Cons_d_mn}) holds, consider the class of functions $\mathcal{F} \equiv  \{f_\theta(x,z) \equiv  (z - \alpha)^2 1(x \le d) + (z - \beta)^2 1(x > d)| \theta = (\alpha,\beta,d) \in [0,1]^3\}$ with the envelope $F(x,z) = 1$. Note that $\mathcal{F}$ is formed by combining three bounded VC classes of functions: $\{(z - \alpha)^2: 0 \leq \alpha \leq 1\}, \{(z - \beta)^2: 0 \leq \beta \leq 1\}$ and $\{1(x \leq d): 0 \leq d \leq 1\}$ through finitely many operations involving addition and multiplication and therefore satisfies the entropy condition in the third display on page 168 of van der Vaart and Wellner (1996). It follows that $\mathcal{F}$ satisfies the conditions of Theorem 2.8.1 of van der Vaart and Wellner (1996) and is therefore uniformly Glivenko-Cantelli for the class of probability measures $\{P_m\}$, i.e.,
\begin{eqnarray}\label{eq:UnifGC_M}
\sup_{m \ge 1} P_m\{\sup_{n \ge k} \|\mathbb{M}_{m,n} - M_m\|_{\infty}
> \epsilon \} \rightarrow 0
\end{eqnarray}
for every $\epsilon > 0$ as $k \rightarrow \infty$, where $\|\cdot\|_\infty$ denotes the supremum (uniform) norm over the function class.
\newline
\newline
Fix $\epsilon >0$ and consider the event $\{\|\hat \theta_{m,n} - \theta_m\| > \epsilon\}$. Since $\theta_m$ minimizes $M_m$ and $\hat{\theta}_{m,n}$ minimizes $\mathbb{M}_{m,n}$, by arguments analogous to the \emph{generic steps} in the proof of Theorem \ref{uniform-cts-mapping}, we have:
\begin{equation}\label{eq:BoundArgMin}
\|\hat \theta_{m,n} - \theta_m\| > \epsilon \Rightarrow \|\mathbb{M}_{m,n} - M_m\|_{\infty} \ge \eta_m(\epsilon)/2\,,
\end{equation}
where
\[ \eta_m(\epsilon) = \inf_{\theta \in [\theta_m - \epsilon \mathbf{1}, \theta_m + \epsilon \mathbf{1}]^c} \{ M_{m}(\theta) - M_m(\theta_m) \,.\]
and $\mathbf{1} = (1,1,1)'$.
\newline
\newline
{\bf Claim:} There exists $\eta > 0$ and an integer $M_0$ such that $\eta_m(\epsilon) \ge \eta > 0$ for all $m \geq M_0$.

We assume the claim for the time being, which yields,
\begin{eqnarray}
P_m\{\sup_{n \ge k} \|\hat \theta_{m,n} - \theta_m \| > \epsilon\} & \le & P_m\{\sup_{n \ge k} \| \mathbb{M}_{m,n} - M_m\|_{\infty} > \eta/2\}, \forall m \geq M_0 \nonumber
\end{eqnarray}
and thus by (\ref{eq:UnifGC_M}), $\sup_{m \geq M_0}\,P\{\sup_{n \ge k} \|\hat \theta_{m,n} - \theta_m \| > \epsilon\} \rightarrow 0 \mbox{ as } k \rightarrow \infty\,$.
This completes the proof of the theorem. \hfill $\Box$ \newline
\newline
{\bf Proof of the Claim:}
Let us bound $M_{m}(\theta) - M_m(\theta_m)$ below as
\begin{eqnarray}
    M_{m}(\theta) - M_m(\theta_m) & = & (M_{m} - M)(\theta) - (M_{m} - M)(\theta_m) + \{M(\theta) - M(\theta_m)\} \nonumber \\
    & \ge & -2 \|M_{m} - M\|_{\infty} + \{M(\theta) - M(\theta_m)\} \nonumber
\end{eqnarray}
As $\|M_{m} - M\|_{\infty} \rightarrow 0$ as $m \rightarrow \infty$, it is enough to show that there exists $\eta > 0$ such that for all sufficiently large $m$, $\inf_{\theta \in [\theta_m - \epsilon \mathbf{1}, \theta_m + \epsilon \mathbf{1}]^c} \{ M(\theta) - M(\theta_m) \} > \eta$. Note that the main difficulty arises because $M(\cdot)$ is not twice-differentiable at $(1/2,0,d^0)$.

We split $M(\theta) - M(\theta_m)$ into two parts as $\left\{ M(\theta) - M \left(1/2,0, d^0\right) \right\} + \left\{ M \left(1/2,0, d^0\right) - M(\theta_m) \right\}$. Notice that by the continuity of $M(\cdot)$, the second term goes to $0$. To handle the first term notice that $M(\theta) - M(1/2,0,d^0) = \int_0^d \{\nu(x) - \alpha\}^2 p_X(x) dx + \int_d^1 \{\nu(x) - \beta\}^2 p_X(x) dx$.

There exists $M_0 \in \mathbb{N}$ such that for all $m > M_0$, we have $\theta_m \in [\theta_0 - (\epsilon/2) \mathbf{1}, \theta_0 + (\epsilon/2) \mathbf{1}]$ as $\theta_m \rightarrow \theta_0$. Observe that for $\theta =(\alpha,\beta,d) \in [\theta_m - \epsilon \mathbf{1}, \theta_m + \epsilon \mathbf{1}]^c$, $m > M_0$, and $d \ge d^0$, we have
\begin{eqnarray}
& & M(\theta) - M(1/2,0,d^0) = \int_0^{d^0} \left( \frac{1}{2} - \alpha \right)^2 p_X(x) dx + \int_{d^0}^d \beta^2 p_X(x) dx  \nonumber \\
& & \qquad \qquad + \int_d^1 \beta^2 p_X(x) dx \qquad \ge \int_0^{d^0} \left( \frac{1}{2} - \alpha \right)^2 p_X(x) dx \ge \left( \frac{\epsilon}{2} \right)^2 \kappa l \nonumber
\end{eqnarray}
as $|1/2 - \alpha| \ge |\alpha_m - \alpha| - |\alpha_m - 1/2| \ge \epsilon/2$. Similarly, for $\theta \in [\theta_m - \epsilon \mathbf{1}, \theta_m + \epsilon \mathbf{1}]^c$, $m > M_0$, and $d < d^0$, we have
\begin{eqnarray}
& & M(\theta) - M(1/2,0,d^0) = \int_0^{d} \left(\frac{1}{2} - \alpha \right)^2 p_X(x) dx + \int_{d}^{d^0} \left( \frac{1}{2} - \beta \right)^2 p_X(x) dx  \nonumber \\
& & \qquad \qquad + \int_{d^0}^1 \beta^2 p_X(x) dx \qquad \ge \int_{d^0}^1 \beta^2 p_X(x) dx \ge \left( \frac{\epsilon}{2} \right)^2 \kappa l  \nonumber
\end{eqnarray}
as $|\beta| \ge |\beta_m - \beta| - |\beta_m| \ge \epsilon/2$. Take $\eta = \kappa l \epsilon^2/4$. This completes the proof of the claim.
\newline
\newline
{\bf Proof of Theorem \ref{tau-cons-theorem}:} We start with some notation. Let $W_m^{(i)} \equiv \sqrt{m} \overline{\epsilon}_{i,\cdot}/ \sigma_0 , i = 1, 2, \ldots, n$, and consider our i.i.d. ``data'' as: $\{X_i,W_m^{(i)}\}_{i=1}^n$. Note that $W_m^{(i)}$ has density $\phi_m(\cdot)$. Let $\mathbb{P}_{n,m}(\cdot)$ denote the empirical measure of these observables and $P_m$ the joint law of $(X_1, W_m^{(1)})$. Let $\sigma_0$ denote the true variance of $\epsilon_{ij}$, and let $\sigma$ denote any such generic value. For a fixed $\sigma >0$ and $h \in \mathbb{R}$
define (with a slight abuse of notation):
\begin{eqnarray}
Z_{im}^\sigma(h) = 1 - \Phi \left( \frac{\sqrt{m}(\overline{Y}_{i \cdot} - \tau_0) - h}{\sigma} \right) & = & 1 - \Phi \left(\frac{\sqrt{m}(\mu(X_i) - \tau_0) - h + \sqrt{m} \overline{\epsilon}_{i,\cdot}}{\sigma} \right), \nonumber \\
\mathbb{M}_{n,m}^\sigma(h) = \frac{1}{n} \sum_{i=1}^n \left\{ Z_{im}^\sigma(h) - \frac{1}{2} \right\}^2 & = & \mathbb{P}_{n,m}\,\left[ Z_{1m}^\sigma(h) - \frac{1}{2} \right]^2,\nonumber
\end{eqnarray}
and note that $\hat{h}_{m,n}^{\hat \sigma} = \arg \min_{h}\,\mathbb{M}_{n,m}^{\hat \sigma}(h) \equiv \sqrt{m}(\hat{\tau}_{m,n}^{\hat \sigma} - \tau_0) $, where $\hat \sigma = \hat \sigma_{n,m}$. Let $h_m^\sigma = \arg \min_{h} M_m^\sigma(h)$ where
\begin{eqnarray}
M_m^\sigma(h) & = & P_m \left[ \frac{1}{2} - \Phi \left( \frac{\sqrt{m}(\mu(X_1) - \tau_0) - h + \sigma_0 W_m^{(1)}}{\sigma} \right) \right]^2 \nonumber \\
& = & \int_0^1 \left[\int_{-\infty}^{\infty} \left\{ \frac{1}{2} - g_m^{\sigma, h}(x,y) \right\}^2 \phi_m(y) dy \right] \, p_X(x) dx \nonumber
\end{eqnarray}
with $g_m^{\sigma, h}(x,y) = \Phi\left(\frac{\sqrt{m}(\mu(x) - \tau_0) - h + \sigma_0 y}{\sigma} \right)$.
\newline
\newline
Let $\epsilon, \xi > 0$ be given. Letting $P_m^n$ denote the $n$-fold product measure of $P_m$, we want to show that $P_m^n\{|\hat h_{m,n}^{\hat \sigma} - 0| > \epsilon\} \le \xi$ for all large $m$ and $n$. Subsequently, we will denote $P_m^n$ by $P_m$ (again in an abuse of notation), but it will be clear from the context whether we are alluding to the product measure. We bound the quantity of interest as
\begin{eqnarray}\label{eq:ReqProb}
P_m\{|\hat h_{m,n}^{\hat \sigma} - 0| > \epsilon\} \le  P_m\{ |\hat h_{m,n}^{\hat \sigma} - h_m^{\hat \sigma} | > \epsilon/2 \} + P_m\{ |h_m^{\hat \sigma} - 0| > \epsilon/2 \}.
\end{eqnarray}
We employ the following steps to complete the proof of the theorem: \newline

\noindent {\it Step 1}: Establish that there exists $\delta_0 > 0 $ and $M_0 > 0$ such that $|\sigma - \sigma_0| \le \delta_0$ and $m \ge M_0$ implies $|h^\sigma_m - 0| < \epsilon/2$.

Notice that as $\hat \sigma$ is a consistent estimator of $\sigma_0$, there exists $M_1$ such that for all $m,n \ge M_1 > 0$, $P_m \{|\hat \sigma_{m,n} - \sigma_0 | \le \delta_0\} \ge 1 - \xi/3$. Therefore, using {\it Step 1}, $P_m \{|h^{\hat \sigma}_m - 0| > \epsilon/2\} \le \xi/3$ for $m \ge \max \{M_0,M_1\}$. \newline

\noindent {\it Step 2}: Note that the first term on the right side in (\ref{eq:ReqProb}) is bounded by
\begin{eqnarray}\label{eq:ReqProb2}
& & P_m\{|\hat h_{m,n}^{\hat \sigma} - h_m^{\hat \sigma}| > \epsilon/2, |\hat \sigma - \sigma_0| \le \delta_0 \} +  P_m\{|\hat \sigma - \sigma_0| > \delta_0 \} \nonumber \\
& \le & P_m \left\{ \sup_{|\sigma - \sigma_0 | \le \delta_0} |\hat h_{m,n}^{\sigma} - h_m^{\sigma}| > \epsilon/2 \right\} + \xi/3
\end{eqnarray}
for all $n,m \ge \max\{M_0,M_1\}$.  Therefore, it is enough to show that for some $M$ (possibly depending on $\epsilon$),
\begin{eqnarray}\label{eq:Cons_Tau_mn}
\sup_{m \ge M} P_m \left\{ \sup_{|\sigma - \sigma_0 | \le \delta_0} | \hat h_{m,n}^\sigma - h_m^\sigma | > \epsilon/2 \right\} \rightarrow 0, \mbox{ as } n \rightarrow \infty.
\end{eqnarray}

\noindent {\it Proof of Step 1}:  We study the behavior of $M_m^\sigma(h)$ as $m \rightarrow \infty$. Note that $g_m^{\sigma,h}(x,y) \rightarrow \Phi(\frac{- h + \sigma_0 y}{\sigma})$, if $x \le d^0$, and $1$ if $x > d^0$, as $m \rightarrow \infty$. Therefore, $M_m^\sigma(h)$ converges point--wise (by an application of the dominated convergence theorem (DCT) along with Scheffe's theorem) to $M^\sigma(h)$, where
\begin{equation}\label{eq:M_sigma}
M^\sigma(h) = c_1^{\sigma}(h) \int_0^{d^0} p_X(x) dx + \frac{1}{4} \int_{d^0}^1 p_X(x) dx < 1/4.
\end{equation}
where $c_1^{\sigma}(h) = \int_{-\infty}^{\infty} \left\{ 1/2 - \Phi\left((-h + \sigma_0 y)/\sigma \right) \right\}^2 \phi(y) dy$. To see this, observe that \\ $\int_{-\infty}^{\infty} \left\{ 1/2 - g_m^{\sigma,h}(x,y) \right\}^2 \phi_m(y) dy$, which is uniformly bounded by a positive constant for all $m,x$, can be decomposed as
\begin{eqnarray}\label{eq:Sheffe}
     \int_{-\infty}^{\infty} \left\{ 1/2 - g_m^{\sigma,h}(x,y) \right\}^2 \phi(y) dy +  \int_{-\infty}^{\infty} \left\{ 1/2 - g_m^{\sigma,h}(x,y) \right\}^2 \{\phi_m(y) - \phi(y)\} dy
\end{eqnarray}
where the first term converges to $c_1(h)$ for $x \le d^0$ and to $1/4$ for $x > d^0$. The second term in (\ref{eq:Sheffe}) converges to 0 by Scheffe's theorem for all $x \in [0,1]$. The convergence of $M_m^\sigma(h)$ now directly follows from  the DCT. Let $h^\sigma = \arg \min M^\sigma(h)$ for $h \in \mathbb{R}$.
\newline
\newline
{\bf Claim 1:} There exists $\delta'> 0$, such that $\sup_{|\sigma - \sigma_0 | \le \delta'} \sup_{h \in \mathbb{R}} |M_m^\sigma(h) - M^\sigma(h)| \rightarrow 0$ as $m \rightarrow \infty$.
\newline
{\bf Proof of Claim 1:}  For notational convenience we will write $\Phi(x)\,(1-\Phi(x))$ as $\Phi\,(1-\Phi)(x)$ in what follows. Choose $\delta'$ such that $0 <\delta' < \sigma_0$. Let $\eta > 0$ be given. Note that
\begin{eqnarray}\label{eq:M_m-M}
M_m^\sigma(h) - M^\sigma(h) = A_m^{\sigma,h} \int_0^{d^0} p_X(x) dx + \int_{d^0}^1 B_m^{\sigma,h}(x) p_X(x) dx \\
\mbox{ where } A_m^{\sigma,h} = \int_{-\infty}^{\infty} \left\{\frac{1}{2} - \Phi \left( \frac{- h + \sigma_0 y}{\sigma} \right) \right\}^2 (\phi_m - \phi)(y) dy \;\; \mbox{ and }\nonumber \\
B_m^{\sigma,h}(x) = \int_{-\infty}^{\infty} \left\{\frac{1}{2} - g_m^{\sigma,h}(x,y) \right\}^2 \phi_m(y) dy  - \frac{1}{4}. \nonumber
\end{eqnarray}
To simplify notation, denote the set $\{(\sigma,h): |\sigma - \sigma_0| \leq \delta', h \in \mathbb{R}\}$ by $\mathcal{C}$. Then,
\begin{eqnarray*}
\sup_{\mathcal{C}}|M_m^{\sigma}(h) - M^{\sigma}(h)| &\leq& F_X(d^0)\,\sup_{\mathcal{C}}\,|A_m^{\sigma,h}| + \sup_{\mathcal{C}}\,\int_{d^0}^1\,
|B_m^{\sigma,h}(x)|\,p_X(x)\,dx \,.
\end{eqnarray*}
Now, $\sup_{\mathcal{C}}\,|A_m^{\sigma,h}| \le  \int_{-\infty}^{\infty} |\phi_m - \phi|(y) dy \rightarrow 0$ by Scheffe's theorem, and
\begin{eqnarray}\label{eq:B_m}
|B_m^{\sigma,h}(x)| & \le & \int_{-\infty}^{\infty} |\phi_m - \phi|(y) dy + \sup_{\mathcal{C}}\,\left| \int_{-\infty}^{\infty} \left\{\frac{1}{2} - g_m^{\sigma,h}(x,y) \right\}^2 \phi(y) dy  - \frac{1}{4} \right| \nonumber \\
& = & o(1) + \sup_{\mathcal{C}} \int_{-\infty}^{\infty} \Phi \,(1 - \Phi) \left(\frac{\sqrt{m}(\mu(x) - \tau_0) - h + \sigma_0 y}{\sigma} \right) \phi(y) dy \nonumber\,.
\end{eqnarray}
Now,
\[\sup_{\mathcal{C}}\,\int_{d^0}^1\,|B_m^{\sigma,h}(x)|\,p_X(x)\,dx = \left(\sup_{\mathcal{C}_{\leq 0}}\, \int_{d^0}^1 \,|B_m^{\sigma,h}(x)|\,p_X(x)\,dx \right) \vee \left(\sup_{\mathcal{C}_{>0}}\,\int_{d^0}^1\,|B_m^{\sigma,h}(x)|\,p_X(x)\,dx \right),\] where $\mathcal{C}_{\leq 0}$ and $\mathcal{C}_{> 0}$ are defined analogously to $\mathcal{C}$, but with $h$ varying over $(-\infty,0]$ and $(0,\infty)$, respectively. Since, for each $x > d^0$, $\sup_{\mathcal{C}_{\leq 0}}\,\int_{-\infty}^{\infty} \Phi \,(1 - \Phi) \left(\frac{\sqrt{m}(\mu(x) - \tau_0) - h + \sigma_0 y}{\sigma} \right) \phi(y) dy$ is easily seen to be dominated by $\sup_{|\sigma - \sigma_0| \leq \delta'}\, \int_{-\infty}^{\infty} (1 - \Phi) \left(\frac{\sqrt{m}(\mu(x) - \tau_0) + \sigma_0 y}{\sigma} \right) \phi(y) dy$
which goes to 0 as $m \rightarrow \infty$, it follows readily that the first term on the right side of the last display is $o(1)$. It remains to deal with the second. To this end, for $\lambda, h > 0$, define $D_m^{\lambda,h} = \{d^0 < x \le 1: |\mu(x) - (\tau_0 + h/\sqrt{m})| \le \lambda\}$. Given $\eta > 0$, there exists $\lambda \equiv \lambda(\eta) > 0$ (but not depending on $h > 0$) such that $\int_{D_m^{\lambda,h}}\,p_X(x)\,dx < \eta$ by Assumption (A) of Theorem \ref{tau-cons-theorem}. Then,
\begin{eqnarray*}
\sup_{\mathcal{C}_{>0}}\,\left| \int_{d^0}^1 B_m^{\sigma,h}(x) p_X(x) dx \right| \le \sup_{\mathcal{C}}\,\left| \int_{D_m^{\lambda,h}} B_m^{\sigma,h}(x) p_X(x) dx \right| + \sup_{\mathcal{C}_{>0}}\,\left| \int_{[d^0,1]-D_m^{\lambda,h}} B_m^{\sigma,h}(x) p_X(x) dx \right|  \\
\le \eta + o(1) + \sup_{\mathcal{C}_{>0}}\,\int_{[d^0,1]-D_m^{\lambda,h}} \int_{-\infty}^{\infty} \Phi (1 - \Phi) \left(\frac{\sqrt{m}(\mu(x) - \tau_0 - h {m^{-1/2}}) + \sigma_0 y}{\sigma} \right) \phi(y) dy \; p_X(x) dx \,.
\end{eqnarray*}
The last term in the above display is readily seen to be bounded by $$ \sup_{\mathcal{C}_{>0}} \,\int_{[d^0,1] - D_m^{\lambda,h}} \int_{-\infty}^{\infty} \max \left\{ \Phi \left(\frac{-\sqrt{m} \lambda + \sigma_0 y}{\sigma} \right),  (1 - \Phi) \left(\frac{\sqrt{m} \lambda + \sigma_0 y}{\sigma} \right) \right\} \phi(y) dy \; p_X(x) dx$$
which, in turn, is no larger than
\[ \int_{[d^0,1]} \int_{-\infty}^{\infty} \sup_{\sigma \in [\sigma_0 - \delta', \sigma_0 + \delta']}\,\max \left\{ \Phi \left(\frac{-\sqrt{m} \lambda + \sigma_0 y}{\sigma} \right),  (1 - \Phi) \left(\frac{\sqrt{m} \lambda + \sigma_0 y}{\sigma} \right) \right\} \phi(y) dy \; p_X(x) dx\]
and this can be made less than $\eta$ for sufficiently large $m$. It follows that \\ $\sup_{\mathcal{C}_{>0}}\,\int_{d^0}^1\,|B_m^{\sigma,h}(x)|\,p_X(x)\,dx
< 3\,\eta$ for all sufficiently large $m$ and Claim 1 follows.
\newline
\newline
{\bf Claim 2:} There exists there exists $\delta_0 > 0 $ and $M_0 > 0$ such that for all $\sigma$ with $|\sigma - \sigma_0 | \le \delta_0$ and $m \ge M_0$, $|h^\sigma_m - 0| < \epsilon/2$.
\newline
\newline
{\bf Proof of Claim 2:} This will be proved by a direct application of Theorem \ref{uniform-cts-mapping}. In that theorem, take $n$ to be $m$, $\mathcal{T}$ to be the set $|\sigma - \sigma_0| \leq \delta'$ and $\mathcal{H}$ to be $\mathbb{R}$. Also, $\mathbb{M}_n^{\tau}$ is now $M_m^{\sigma}$ and $M^{\tau}$ is now $M^{\sigma}$. We will show that $M^{\sigma}$ is uniquely minimized at a point, say $h^{\sigma}$, and also that $\inf_{|\sigma - \sigma_0| \leq \delta'}\,\inf_{|h-h^{\sigma}| > \eta}(M^{\sigma}(h) - M^{\sigma}(h^{\sigma})) > 0$ for every $\eta > 0$, whence, by Claim 1, it will follow that $\sup_{|\sigma - \sigma_0| \leq \delta'}\,|h_m^{\sigma} - h^{\sigma}|$ converges to 0 with increasing $m$. But, as will also be seen, $h^{\sigma}$ equals 0 for all $\sigma$ and hence Claim 2 follows with $\delta_0$ taken to be $\delta'$.
\newline
\newline
From the form of $M^{\sigma}(h)$ (see \ref{eq:M_sigma}) it suffices to show that $\inf_{|\sigma - \sigma_0| \leq \delta'}\,\inf_{|h-h^{\sigma}| > \eta}(c_1^{\sigma}(h) - c_1^{\sigma}(h^{\sigma})) > 0$, where $h^{\sigma}$ is the unique point at which $c_1^{\sigma}$ is minimized. We now make some change of variables to facilitate the ensuing argument. Define $\lambda = \sigma/\sigma_0$ and $s = h/\sigma_0$. Then $|\sigma - \sigma_0| \leq \delta' \Leftrightarrow |\lambda - 1| \leq \delta^{''}$ (for some $\delta^{''} < 1$) and $\Phi((-h + \sigma_0\,y)/\sigma) = \Phi(\lambda^{-1}(y-s))$. Defining
\[ \tilde{c}_1^{\lambda}(s) = \int_{-\infty}^{\infty}\, \left[\frac{1}{2} - \phi(\lambda^{-1}(y-s)) \right]^2\,\phi(y)\,dy \,,\] it suffices to show that $\inf_{|\lambda-1|\leq\delta^{''}}\,\inf_{|s-s^{\lambda}|\geq \eta/\sigma_0}\,(\tilde{c}_1^{\lambda}(s) - \tilde{c}_1^{\lambda} (s_{\lambda})) > 0$ where $s^{\lambda}$ is the unique minimizer of $\tilde{c}_1^{\lambda}$. It is easy to see that $\tilde{c}_1^{\lambda}(s) = E\,\left[\frac{1}{2} - \Phi(\lambda^{-1}(Z-s))\right]^2$ where $Z$ is a standard normal random variable. By the symmetry of $Z$ about 0, it follows easily that $\tilde{c}_1^{\lambda}(s) = \tilde{c}_1^{\lambda}(-s)$. Furthermore $\tilde{c}_1^{ \lambda}(s)$ is strictly increasing for $s > 0$, and is therefore strictly decreasing for $s \leq 0$, showing that $0$ is the unique minimizer of $\tilde{c}_1^{\lambda}$. Hence $s_{\lambda} = 0$ for all $\lambda$, showing that $h^{\sigma} = 0$ for all $\sigma$. Thus, \[ \inf_{|\lambda-1|\leq\delta^{''}}\,\inf_{|s-s^{\lambda}|\geq \eta/\sigma_0}\,(\tilde{c}_1^{\lambda}(s) - \tilde{c}_1^{\lambda} (s_{\lambda})) = \inf_{|\lambda-1|\leq \delta^{''}}(\tilde{c}_1^{\lambda}(\eta/\sigma_0) - \tilde{c}_1^{\lambda}(0)) \,.\]
Since $\tilde{c}_1^{\lambda}(\eta/\sigma_0) - \tilde{c}_1^{\lambda}(0)$ is continuous and positive for each $\lambda$, its infimum on the set $|\lambda - 1| \leq \delta^{''}$, which must be achieved, is positive.
\newline
\newline
{\it Proof of Step 2}: Consider the class of functions $\mathcal{F}_{\infty} \equiv \cup_{m}\,\mathcal{F}_m$ where
$\mathcal{F}_m \equiv  \{f_{h,\sigma}(x,w) \equiv  \{1/2 - \Phi(\sqrt{m}(\mu(x) - \tau_0)/\sigma + h/\sigma + w \frac{\sigma_0}{\sigma})\}^2 | \tau \in \mathbb{R}, \sigma \in [\sigma_0 - \delta_0, \sigma_0 + \delta_0]\}$. This is a subclass of the large class of functions $\mathcal{G} = \{g_{\alpha,\beta,\gamma}(x,w) \equiv  [1/2 - \Phi(\alpha \mu(x) +  \beta w + \gamma)]^2 | (\alpha,\beta,\gamma) \in \mathbb{R}^3\}$. Note that the class $\{\alpha \mu(x) +  \beta w + \gamma\}$ as $(\alpha,\beta,\gamma)$ varies in $\mathbb{R}^3$ forms a finite dimensional vector space of measurable functions and is therefore VC. Hence, $1/2 - \Phi(\alpha \mu(x) +  \beta w + \gamma)$, being a bounded monotone transformation of a VC class of functions, is bounded VC and consequently, so is the class $\mathcal{F}_{\infty}$. Thus, $\mathcal{F}_{\infty}$ satisfies the entropy condition in the third display on Page 168 of van der Vaart and Wellner (1996) and therefore the conditions of Theorem 2.8.1 of van der Vaart and Wellner (1996) and is uniformly Glivenko--Cantelli for the class of probability measures $\{P_m\}$, i.e., for any given $\zeta > 0$,
\[ \sup_{m \ge 1} P_m\{\sup_{k \ge n} \| \mathbb{M}_{m,k}^\sigma - M_m^\sigma \|_{\mathcal{F}_{\infty} > \zeta} \} \rightarrow 0 \mbox{ as } n \rightarrow \infty \]
and therefore,
\begin{eqnarray}\label{eq:UnifGC_MTau}
\sup_{m \ge 1} P_m\{\sup_{k \ge n} \|\mathbb{M}_{m,k}^\sigma - M_m^\sigma\|_{\mathcal{F}_m} > \zeta \} \rightarrow 0 \;\;\mbox{as} \; n \rightarrow \infty.
\end{eqnarray}
Next, using techniques similar to that from proving (\ref{eq:BoundArgMin}), we can show that \begin{equation}\label{final-implication}
\sup_{|\sigma - \sigma_0| \le \delta_0} |\hat h_{m,n}^{\sigma} - h_m^{\sigma}| > \epsilon/2 \Rightarrow \|\mathbb{M}_{m,n}^\sigma - M_m^\sigma\|_{\mathcal{F}_m} \ge \eta_m(\epsilon/2) \end{equation}
where $\eta_m (\epsilon) = \inf_{|\sigma - \sigma_0| \le \delta_0} \inf_{|h  - h_m^\sigma| > \epsilon/2} \{ M_{m}^\sigma(h) - M_m^\sigma(h_m^\sigma) \}$.

{\bf Claim 3:} There exists $\eta > 0$ and an integer $\tilde{M}$ such that $\eta_m(\epsilon) \ge \eta > 0$ for all $m \ge \tilde{M}$.

{\bf Proof of Claim 3:}
By Claim 2, for all sufficiently large $m$, uniformly for $\sigma \in [\sigma_0 - \delta_0, \sigma_0 + \delta_0]$, we have $[h_m^{\sigma} - \epsilon/2, h_m^{\sigma} + \epsilon/2]^c \subset [- \epsilon/4, \epsilon/4]^c$. We conclude, that for all sufficiently large $m$, \[ \eta_m(\epsilon) \geq  \tilde{\eta}_m(\epsilon) \equiv \inf_{|\sigma - \sigma_0| \le \delta_0} \inf_{|h - 0| > \epsilon/4} \{ M_{m}^\sigma(h) - M_m^\sigma(h_m^\sigma) \} \,.\] For $h$ and $\sigma$ such that $|h - 0|> \epsilon/4$ and $|\sigma - \sigma_0 | \le \delta_0$, we can bound $M_{m}^\sigma(h) - M_m^\sigma(h_m^\sigma)$ below as
\begin{eqnarray}
    M_{m}^\sigma(h) - M_m^\sigma(h_m^\sigma) & = & (M_{m}^\sigma - M^\sigma)(h) - M_{m}^\sigma (h_m^\sigma) + M^{\sigma}(h) \nonumber \\
    & \ge & - \sup_{|h - 0| > \epsilon/4} |(M_{m}^\sigma - M^\sigma)(h)| - M_m^\sigma(0) + M^\sigma(h) \nonumber \\
    & \ge & - \sup_{|\sigma - \sigma_0 | \le \delta_0} \sup_{|h - 0| > \epsilon/4} |(M_{m}^\sigma - M^\sigma)(h)| \nonumber \\
    & - & \sup_{|\sigma - \sigma_0 | \le \delta_0} |(M_m^\sigma - M^\sigma)(0)| \nonumber \\
    &+ & \inf_{|\sigma - \sigma_0 | \le \delta_0} \inf_{|h - 0| > \epsilon/4} [M^\sigma(h) - M^\sigma(0)] \nonumber
\end{eqnarray}
As $\sup_{|\sigma - \sigma_0 | \le \delta_0} \sup_{|h - 0| > \epsilon/4} |(M_{m}^\sigma - M^\sigma)(h)| \rightarrow 0$ and $\sup_{|\sigma - \sigma_0 | \le \delta_0} |(M_m^\sigma - M^\sigma)(0)| \rightarrow 0$ as $m \rightarrow \infty$, and $ \inf_{|\sigma - \sigma_0 | \le \delta_0} \inf_{|h - 0| > \epsilon/4} [M^\sigma(h) - M^\sigma(0)]  =: 2\,\eta > 0$, it follows that for all large $m$, $\tilde{\eta}_m(\epsilon) \ge \eta > 0$; therefore, for all sufficiently large $m$, say $m \geq \tilde{M}$, $\eta_m(\epsilon) \ge \eta >0$. This completes the proof of the claim.
\newline
\newline
Hence, for all $m \geq \tilde{M}$,
\begin{eqnarray}
 \sup_{m \geq \tilde{M}}\,P_m\left\{\sup_{|\sigma - \sigma_0 | \le \delta_0}\,| \hat{h}_{m,n}^{\sigma} - h_m^{\sigma} | > \epsilon/2 \right\} &\le&
\sup_{m \geq \tilde{M}}P_m \left\{\sup_{k \geq n} \|\mathbb{M}_{m,n}^{\sigma} - M_m^{\sigma} \|_{\mathcal{F}_m} > \eta_m(\epsilon)/2\right\} \nonumber \\
\qquad \qquad \qquad \qquad \qquad \qquad \qquad \qquad &\le& \sup_{m \geq \tilde{M}}P_m \left\{\sup_{k \geq n}\,\|\mathbb{M}_{m,k}^{\sigma} - M_m^{\sigma} \|_{\mathcal{F}_m} > \eta/2 \right\} \nonumber
\end{eqnarray}
and (\ref{eq:Cons_Tau_mn}) follows from (\ref{eq:UnifGC_MTau}).  \hfill $\Box$
\newline
\newline
{\bf Proof of Theorem \ref{homosced-err-pvalcons}:} For notational simplicity, we refer to $\hat{d}_{m,n}$ and $\hat{\sigma}_{m,n}$ in this proof as $\hat{d}$ and $\hat{\sigma}$ respectively. We borrow notation from the proof of Theorem \ref{tau-cons-theorem}, but note that now $\sigma$ is a constant (as opposed to being a function). We seek to show that for given $\epsilon > 0$, $P_m\{|\hat{d} - d^0 | > \epsilon\} \rightarrow 0$, as $m,n \rightarrow \infty$. Define: \[ m_d^\sigma(Z_{1m},X_1) = \{Z_{1m}^{\sigma}(\tau_0) - 1/2\}^2 1(X_1 \leq d) + \{Z_{1m}^{\sigma}(\tau_0) - 0\}^2 1(X_1 > d),\] and let $M_m^\sigma(d) = P_m [m_d^\sigma(Z_{1m},X_1)]$ and $\mathbb{M}_{m,n}^\sigma(d) = \mathbb{P}_{n,m} \left[ m_d^\sigma(Z_{1m},X_1)\right]$. Also, define $d_{\sigma}^{m} = \arg \min_{d} M_m^\sigma(d)$ and $\hat{d}_{\sigma}^{m,n} = \arg \min_{d} \mathbb{M}_{m,n}^\sigma(d)$.
\newline
\newline
{\bf Step 0:} Establish that for $m$ sufficiently large, $\sup_{|\sigma - \sigma_0| \le \delta_0} | d_{\sigma}^m - d^0 | \le \epsilon$, for a pre-assigned $\epsilon > 0$, where $\delta_0 > 0$ is a small number chosen to depend on $\epsilon$.
\newline
\newline
This quantity of interest is dominated by:
\begin{equation}\label{eq:Bound_d_hat}
P_m\{| \hat{d} - d_{\hat{\sigma}}^m | > \epsilon/2\} + P_m\{| d^m_{\hat{\sigma}} - d^0 | > \epsilon/2\},
\end{equation}
where $\hat d \equiv \hat{d}_{\hat{\sigma}}^{m,n}$. By Step 0 and the consistency of $\hat{\sigma}$ for $\sigma$, the second term in the above display can be made (arbitrarily) small for all sufficiently large $m$. By the consistency of $\hat{\sigma}$, again, it suffices to show that for some $M$ and $\delta_0 > 0$ chosen appropriately, \[ \mbox{sup}_{m \geq M}\,P_m\,\left\{\sup_{|\sigma - \sigma_0| \le \delta_0}\,| \hat{d}_{\sigma}^{m,n} - d_{\sigma}^m | > \epsilon /2 \right\} \rightarrow 0, \;\mbox{as}\; n \rightarrow \infty \,.\]
The above display is comparable to (\ref{eq:Cons_Tau_mn}) in the proof of Theorem \ref{tau-cons-theorem} and follows via arguments similar (in fact, much simpler) to those following (\ref{eq:Cons_Tau_mn}), involving uniformly Glivenko--Cantelli classes of functions. We omit the details and in the remainder of the proof, focus on establishing Step 0.
\newline
\newline
Recall how $W_m^{(1)}$ was defined in the proof of Theorem \ref{tau-cons-theorem}. We denote the distribution of $W_m^{(1)}$ by $\Phi_m$ and note that $\Phi_m$ converges weakly to $\Phi$. Note that
\begin{eqnarray}\label{eq:M_m^sigma}
M_m^\sigma(d) & = & \int_0^d E \left[\frac{1}{2} - \Phi\left(\frac{\mu(x)-\tau_0}{\sigma m^{-1/2}} + \frac{\sigma_0}{\sigma} W_m^{(1)} \right) \right]^2 p_X(x) dx \nonumber \\
& & \qquad \qquad + \int_d^1 E \left[1 - \Phi\left(\frac{\mu(x)-\tau_0}{\sigma m^{-1/2}} + \frac{\sigma_0}{\sigma} W_m^{(1)} \right)\right]^2 p_X(x) dx
\end{eqnarray}
The point-wise limit of $M_m^\sigma(d)$ as $m \rightarrow \infty$ is given by
\begin{eqnarray}\label{eq:M_sigma_d}
M^\sigma(d) = \left\{
    \begin{array}{ll}
        c_1 \int_0^d p_X(x) dx + c_2 \int_d^{d^0} p_X(x) dx, & \mbox{ for } d \le d^0, \\
        c_1 \int_0^{d^0} p_X(x) dx + 1/4 \int_{d^0}^d p_X(x) dx, & \mbox{ for } d > d^0,
    \end{array}
    \right.
\end{eqnarray}
where $c_1 \equiv \int_{-\infty}^{\infty} \{1/2 - \Phi(\sigma_0 y/\sigma)\}^2 \phi(y) dy < \int_{-\infty}^{\infty} \{1 - \Phi(\sigma_0 y/\sigma)\}^2 \phi(y) dy \equiv c_2$. We first show that
\begin{equation}\label{eq:M_m-M}
\sup_{|\sigma - \sigma_0| \le \delta_0, d \in [0,1]} |M_m^\sigma(d) - M^\sigma(d)| \rightarrow 0
\end{equation}
as $m \rightarrow \infty$ for some $\delta_0 > 0$ sufficiently small.
\newline
\newline
Choose $\delta_0$ such that $\sigma_0 > \delta_0 > 0$. To show that the convergence is uniform in $d \in [0,1]$ and $\sigma \in [\sigma_0 - \delta_0, \sigma_0 + \delta_0]$, we define $g_m(\eta,\lambda) = E(\Phi(\lambda + \eta\,W_m^{(1)}))$, for $\lambda \in \mathbb{R}$ and $\eta > 0$. For each fixed $\eta$, $g_m(\eta,\lambda)$ is continuous in $\lambda$, converges to 0 as $\lambda \rightarrow -\infty$, to 1 as $\lambda \rightarrow \infty$ and is strictly monotone in $\lambda$.
\newline
\newline
Also define $g(\eta,\lambda) = E(\Phi(\lambda + \eta Z))$, where $Z$ is a standard normal random variable. Let $\epsilon > 0$ be given. By Assumption $(b)$ we can get $\beta > 0$ such that $\int_{d^0}^{d^0 + \beta} p_X(x) dx < \epsilon$. Note that by Assumption $(a)$ of the theorem there exists $\eta > 0$ such that $\mu(x)  - \tau_0 > \eta$ for all $x \ge d^0 + \beta$. Let $S_m = \Phi\left(\frac{\sigma_0}{\sigma} W_m^{(1)} \right)$ and $S = \Phi\left(\frac{\sigma_0}{\sigma} Z \right)$, where $Z \sim N(0,1)$. After some simplification, using (\ref{eq:M_m^sigma}) and (\ref{eq:M_sigma_d}), we can bound $|M_m^\sigma(d) - M^\sigma(d)|$ for $d < d^0$ by
\begin{eqnarray}
    \left\{ E \left(1/2 - S_m \right)^2 - E \left(1/2 - S \right)^2 \right\} \int_0^d p_X(x) dx + \left\{ E \left(1 - S_m \right)^2 - E \left(1 - S \right)^2 \right\} \int_{d}^{d^0} p_X(x) dx \nonumber \\
    + \epsilon + E \left[1 - \Phi\left(\frac{\eta}{\sigma m^{-1/2}} + \frac{\sigma_0}{\sigma} W_m^{(1)} \right) \right]^2 \int_{d^0 + \beta}^{1} p_X(x) dx. \qquad \qquad \nonumber
\end{eqnarray}
We can analogously bound $|M_m^\sigma(d) - M^\sigma(d)|$ for $d > d^0$. Therefore, to show that (\ref{eq:M_m-M}) holds, we need to show that
\begin{eqnarray}
    \sup_{|\sigma - \sigma_0| \le \delta_0} \left| E \left[ \Phi\left(\frac{\sigma_0}{\sigma} W_m^{(1)} \right) \right] - E \left[\Phi \left( \frac{\sigma_0}{\sigma} Z \right) \right] \right| & \rightarrow & 0, \;\;\; \mbox{ and }  \label{eq:c1}  \\
    \sup_{|\sigma - \sigma_0| \le \delta_0} \left| E \left[ \Phi\left(\frac{\eta}{\sigma m^{-1/2}} + \frac{\sigma_0}{\sigma} W_m^{(1)} \right) \right] - 1 \right| & \rightarrow & 0, \label{eq:c2}
\end{eqnarray}
as $m \rightarrow \infty$. To show that (\ref{eq:c1}) and (\ref{eq:c2}) hold, we first observe that
\begin{eqnarray}
    |g_m(\eta, \lambda_m) - g(\eta, \lambda)| \le  |g_m(\eta, \lambda_m) - g(\eta, \lambda_m)| +  |g(\eta, \lambda_m) - g(\eta, \lambda)| \nonumber
\end{eqnarray}
where $\lambda_m \rightarrow \lambda$ as $m \rightarrow \infty$, with $\lambda = 0$ or $\lambda = \infty$. The second term on the right side can be controlled by using the continuity of $g(\eta,\lambda)$ for a fixed $\eta$, and the first term can be bounded by noticing that
\begin{eqnarray*}
& & \sup_{\lambda \in \mathbb{R}, \eta \in [1-\kappa_0, 1+\kappa_0]}\,\left|\,g_m(\eta,\lambda) - g(\eta,\lambda) \right| \\
& \leq & E\,\left[\,\sup_{\lambda \in \mathbb{R}, \eta \in [1-\kappa_0, 1+\kappa_0]}\,\left|\Phi(\lambda + \eta\,W_m^{(1)}) - \Phi(\lambda + \eta\,Z)\right| \,\right]  \rightarrow 0  \;\;\mbox{as}\;\; m \rightarrow \infty\,
\end{eqnarray*}
via an application of the DCT, where $W_m^{(1)}$ can be assumed to converge almost surely to $Z$ (using Skorohod embedding). This proves (\ref{eq:M_m-M}).
\newline
\newline
Note that $\arg \min_{d \in [0,1]} M^\sigma(d) = d^0$. Using techniques similar to that used in proving (\ref{eq:BoundArgMin}), we can show that
\begin{equation}\label{eq:dm-d0}
\sup_{|\sigma - \sigma_0| \le \delta_0} | d_{\sigma}^m - d^0 | > \epsilon \; \Rightarrow \; \sup_{|\sigma - \sigma_0| \le \delta_0, d \in [0,1]} |M_m^\sigma(d) - M^\sigma(d)| > \eta(\epsilon)/2
\end{equation}
where $\eta(\epsilon) = \inf_{|\sigma - \sigma_0| \le \delta_0} \inf_{| d_{\sigma}^m - d^0 | > \epsilon} \{M^\sigma(d_{\sigma}^m) - M^\sigma(d^0)\} > 0$ (follows from (\ref{eq:M_sigma_d})). But, as (\ref{eq:M_m-M}) holds, (\ref{eq:dm-d0}) cannot hold for all large $m$, thereby completing the proof of Step 0. \hfill $\Box$
\newline
\newline
{\bf Proof of Theorem \ref{thm:consTauUnk}:} Letting $Z_{im}^{\sigma}(\tau)$ as in (\ref{basic-def-zim}) (in Section \ref{sec:unknown-tau}) and $m_d^{\sigma,\tau}(Z_{1m},X_1) = \{Z_{1m}^\sigma(\tau) - 1/2\}^2 1(X_1 \le d) + \{Z_{1m}^\sigma(\tau) - 0\}^2 1(X_1 > d)$ , define
\begin{eqnarray}\label{eq:M_n_m}
    \mathbb{M}_{m,n}^{\sigma, \tau} (d)  =  \mathbb{P}_{n,m} [m_d^{\sigma,\tau}(Z_{1m},X_1)] \;\; \mbox{ and } \;\;
    {M}_{m}^{\sigma, \tau} (d)  = {P}_{m} [m_d^{\sigma,\tau}(Z_{1m},X_1)].
\end{eqnarray}
Define $\tilde{d}_{\sigma,\tau}^{m,n} = \arg \min_{d \in [0,1]} \mathbb{M}_{m,n}^{\sigma,\tau} (d)$ and $d_m^{\sigma,\tau} = \arg \min_{d \in [0,1]} {M}_{m}^{\sigma, \tau} (d)$ and note that $\tilde{d}_{m,n} =  \tilde{d}_{\tilde{\sigma},\tilde{\tau}}^{m,n}$. Let $\epsilon > 0$ be given. We seek to show that: $P_m(|\tilde{d}_{m,n} - d^0| > \epsilon) \rightarrow 0$ as $m,n \rightarrow \infty$, to which end it suffices to show that both $P_m(|\tilde{d}_{m,n} - d_{\tilde{\sigma},\tau_0}^{m}| > \epsilon/2)$ and $P_m(|d_{\tilde{\sigma},\tau_0}^{m} - d^0| > \epsilon/2)$ go to 0. The second term is easily handled on noting that $\sup_{|\sigma-\sigma_0| \leq \delta_0}|d_{\sigma,\tau_0}^m - d^0| \leq \epsilon/2$ for some appropriately chosen $\delta_0$ (using Step 0 of Theorem \ref{homosced-err-pvalcons}) and the consistency of $\tilde{\sigma}_{m,n}$ for $\sigma_0$. To handle the first term, we show that
\[ P_m \left( \sup_{|\sigma-\sigma_0| \leq \delta_0}\,|\tilde{d}_{\sigma,\tilde \tau}^{m,n} - d_{\sigma,\tau_0}^{m}| > \epsilon/2 \right) \rightarrow 0\,,\]
and this, again in conjunction with the consistency of $\tilde{\sigma}_{m,n}$ implies the convergence of the first term to 0. We next introduce some notation. For a real-valued function $x^{\sigma}$ defined on $[0,1]$, define  $\| x^{\sigma} \| = \sup_{|\sigma - \sigma_0| \le \delta_0} \sup_{d \in [0,1]} |x^{\sigma}(d)|$, where $\sigma_0 > \delta_0 > 0$. By arguments analogous to \emph{generic steps}, we have:
\[ P_m \left( \sup_{|\sigma-\sigma_0| \leq \delta_0}\,|\tilde{d}_{\sigma,\tilde \tau}^{m,n} - d_{\sigma,\tau_0}^{m}| > \epsilon/2 \right) \leq P_m(\|\mathbb{M}_{m,n}^{\sigma,\tilde{\tau}} - M_m^{\sigma,\tau_0}\|\geq \eta_m(\epsilon)/2) \,,\] with $\eta_m(\epsilon) = \inf_{|\sigma-\sigma_0|\leq \delta_0}\,\inf_{|d-d_{\sigma,\tau_0}^m|>\epsilon/2} \{M_m^{\sigma,\tau_0}(d) -
M_m^{\sigma,\tau_0}(d_{\sigma,\tau_0}^m)\}$. By using arguments similar to that of Claim 3 of Theorem \ref{tau-cons-theorem} and the Claim of Theorem \ref{consistency--theorem} (following (\ref{eq:BoundArgMin})), we can show that $\eta_m(\epsilon) > \eta > 0$ for all sufficiently large $m$,
whence it suffices to show that $P_m(\|\mathbb{M}_{m,n}^{\sigma,\tilde{\tau}} - M_m^{\sigma,\tau_0}\|\geq \eta/2)$ goes to 0 as $m,n \rightarrow \infty$. Now,
\begin{eqnarray}\label{eq:Group_MTau_d}
\|\mathbb{M}_{m,n}^{\sigma,\tilde \tau} - M_m^{\sigma,\tau_0}\| \le \|\mathbb{M}_{m,n}^{\sigma,\tilde \tau} - \mathbb{M}_{m,n}^{\sigma,\tau_0}\| + \|\mathbb{M}_{m,n}^{\sigma, \tau_0} - M_m^{\sigma,\tau_0}\| \,.
\end{eqnarray}
Using arguments (involving universal Glivenko-Cantelli classes) similar to the proof of Step 2 in Theorem \ref{tau-cons-theorem}, we can easily show that: $\sup_{m \geq 1}\,P_m(\sup_{k \geq n}\,\|\mathbb{M}_{m,k}^{\sigma,\tau_0} - M_m^{\sigma,\tau_0}\| > \eta/4) \rightarrow 0$
as $n \rightarrow \infty$, whence it readily follows that $P_m(\|\mathbb{M}_{m,n}^{\sigma,\tau_0} - M_m^{\sigma,\tau_0}\| > \eta/4) \rightarrow 0$
as $m,n \rightarrow \infty$. That $P_m(\|\mathbb{M}_{m,n}^{\sigma,\tilde \tau} - \mathbb{M}_{m,n}^{\sigma,\tau_0}\| > \eta/4)$ goes to 0 follows from the fact that $\|\mathbb{M}_{m,n}^{\sigma,\tilde \tau} - \mathbb{M}_{m,n}^{\sigma,\tau_0}\| \leq \tilde{K}\,\sqrt{m}(\tilde{\tau}_{m,n} - \tau_0)$ which goes to
0 by assumption, as $m,n \rightarrow \infty$. This last inequality follows from the fact that for any $d \in [0,1]$:
\begin{eqnarray}
|\mathbb{M}_{m,n}^{\sigma,\tilde \tau}(d) - \mathbb{M}_{m,n}^{\sigma,\tau_0}(d)| & \le & \frac{1}{n} \sum_{i:X_i \le d} 3 |Z_{im}^{\sigma}(\tilde \tau) - Z_{im}^{\sigma}(\tau_0)| + \frac{1}{n} \sum_{i:X_i > d} 2 |Z_{im}^{\sigma}(\tilde \tau) - Z_{im}^{\sigma}(\tau_0)| \nonumber \\
& \le &  \frac{3}{n} \sum_{i=1}^n |Z_{im}^{\sigma}(\tilde \tau) - Z_{im}^{\sigma}(\tau_0)| \le \frac{\sqrt{m} (\tilde \tau - \tau_0)}{\sigma} \frac{1}{n} \sum_{i=1}^n \phi(\xi^*_i)/\sigma. \nonumber \\
& \le & K \frac{\sqrt{m} (\tilde \tau - \tau_0)}{\sigma} \stackrel{p}{\rightarrow} 0
\end{eqnarray}
for some universal constant $K$, whence $\tilde{K}$ can be taken to be $K/(\sigma_0 - \delta_0)$. $\Box$.
\newline
\newline
\noindent
{\bf Justification of the subsampling procedure:} We consider an asymptotic paradigm, where $m$ is viewed as fixed, and $n$ as increasing to infinity in the setting of a \emph{known} $\tau_0$ (which can then be taken to be 0 without loss of generality). Let the setting be that of Theorem  \ref{consistency--theorem} and consider fitting a stump $\xi_d(x) = (1/2)\,1(x \leq d)$ as a \emph{working model} for $\nu_m$. Recall that the underlying setting corresponds to a  regression one with i.i.d. observations $\{X_i, Z_{im}\}_{i=1}^n$. Letting $P_m$ denote the distribution of $(X_1, Z_{1,m})$, the best-fitting stump in the population is characterized by the parameter $d^m : = \arg \min\,M_m(d)$, where $M_m(d) = P_m\,[(Z_{1,m} - (1/2))^2\,1(X_1 \leq d) + Z_{1m}^2\,1(X_1 > d)]$. Setting the derivative of $M_m$ with respect to $d$, to zero, yields the normal equation $\nu_m(d^m) = 1/4$. Under reasonably modest assumptions on the underlying model, $d^m$ is unique and provides an upper bound on $d^0$ and the larger the $m$, the tighter the bound. A level $1-\alpha$ confidence interval for $d^m$ can then be used as a \emph{surrogate} for a level $1-\alpha$ confidence interval for $d^0$. Letting $\hat{d}_{m,n}$ be the least squares estimate of $d^0$ obtained by minimizing $\sum_{i=1}^n\,[(Z_{i,m} - 0.5)^2\,1(X_i \leq d) + Z_{i,m}^2\,1(X_i > d)]$ over all $d$, it can be shown, by adapting the techniques of Banerjee and McKeague (2007), that for a fixed $m$, $n^{1/3}\,(\hat{d}_{m,n} - d^m)$ converges to a continuous symmetric but non-Gaussian distribution, namely Chernoff's distribution (studied in Groeneboom and Wellner (2001)), as $n \rightarrow \infty$. As this distribution depends upon hard to estimate nuisance parameters in the model, subsampling without replacement (as in Politis, Romano and Wolf (1999)) or the $m$ out of $n$ bootstrap can be used to extract asymptotic confidence intervals for $d^m$. Owing to the non-standard nature of the asymptotics involved, the standard Efron-type bootstrap fails in this situation (see Sen, Banerjee and Woodroofe (2010) and references therein).
\newline
\newline
Note that the implemented subsampling procedure, while relying on the above results in spirit, does take some liberties in its implementation. Firstly, the Method 1 based confidence intervals require estimation of $\tau$ since it is unknown for our application. Secondly, the theoretical results above are not immediately applicable to the confidence intervals using Method 2 which involves a non-trivial modification of the $p$--values used in Method 1. Nevertheless, it seems reasonable to conjecture the same $n^{1/3}$ rate of convergence to $d^m$ for the least squares estimates obtained by this approach, with a non--degenerate continuous distribution, which would then validate the use of subsampling.

\end{document}